\pdfoutput=1
\documentclass[a4paper,times,3p]{elsarticle}

\usepackage[utf8]{inputenc} 
\usepackage[T1]{fontenc}    
\usepackage{microtype}      
\usepackage[font=footnotesize,labelfont=bf]{caption}    
\usepackage{amssymb} 
\usepackage{graphicx} 
\usepackage{adjustbox}
\usepackage{pgf}
\usepackage{pdfpages}

\usepackage{hyperref}
\usepackage{layouts}
\usepackage{tabularx}
\usepackage{graphbox} 
\usepackage{float}
\usepackage{multirow}
\usepackage{verbatim} 
\usepackage{amsmath} 
\DeclareMathOperator{\Tr}{Tr} 

\usepackage{xcolor}
\usepackage{tikz}   
\usepackage{tikz-3dplot}
\usetikzlibrary{angles}
\usetikzlibrary{calc}
\usetikzlibrary{quotes}

\usepackage{cleveref} 
\usepackage{booktabs} 
\usepackage{array}   
\usepackage{placeins} 

\usepackage[acronym,nowarn,hyperfirst=false]{glossaries}

\newacronym{PDE}{PDE}{partial differential equation}
\newacronym{CFD}{CFD}{computational fluid dynamics}
\newacronym[longplural={degrees of freedom}]
           {DOF}{DOF}{degree of freedom}
\newacronym{MSE}{MSE}{mean-squared error}
\newacronym{RMSE}{RMSE}{root-mean-square error}
\newacronym{ML}{ML}{machine learning}
\newacronym{LHS}{LHS}{latin-hypercube-sampling}
\newacronym{TSE}{TSE}{twin-screw extruder}

\newacronym{FEM}{FEM}{finite element method}
\newacronym{WLF}{WLF}{William-Landel-Ferry}
\newacronym{SRMUM}{SRMUM}{snapping reference mesh update method}
\newacronym{DSD}{DSD}{deforming spatial domain}
\newacronym{SST}{SST}{stabilized space-time}
\newacronym{SUPG}{SUPG}{streamline-upwind/Petrov-Galerkin}
\newacronym{PSPG}{PSPG}{pressure-stabilizing/Petrov-Galerkin}
\newacronym{LBB}{LBB}{Ladyzhenskaya-Babu\v{s}ka-Brezzi}

\newacronym{ROM}{ROM}{reduced order modeling}
\newacronym{POD}{POD}{proper orthogonal decomposition}
\newacronym{SVD}{SVD}{singular value decomposition}
\newacronym{EIM}{EIM}{empirical interpolation method}
\newacronym{TPWL}{TPWL}{trajectory piece-wise linear}
\newacronym{SEY}{SEY}{Schimdt-Eckart-Young}

\newacronym{ANN}{ANN}{artificial neural network}
\newacronym{FNN}{FNN}{feedforward neural network}
\newacronym{SGD}{SGD}{stochastic gradient descent}
\newacronym{NAG}{NAG}{Nesterov accelerated gradient}
\newacronym{GPR}{GPR}{Gaussian process regression}
\newacronym{GP}{GP}{Gaussian process}
\newacronym{RBF}{RBF}{radial basis function}
\newacronym{L-BFGS}{L-BFGS}{limited-memory Broyden-Fletcher-Goldfarb-Shanno}

\newcommand{\matr}[1]{\mbox{\boldmath{$\mathsf {#1}$}}}
\newcommand{\vekt}[1]{\mbox{$\boldsymbol{#1}$}}

\newcommand{\scaleNamedInferno}[3]
{
\def\colorscale{figures/colorscale_inferno.png}
\begin{tikzpicture}
      \node[anchor=south west,inner sep=0] at (0,0)
      {\includegraphics[width=3cm, height=0.3cm]{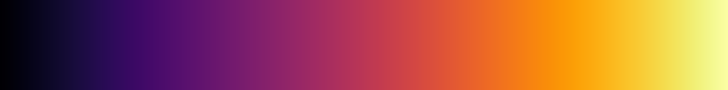}};
      \node (h) at (0,-0.3) {#1};
      \node (h) at (1.5,-0.3){#3};
      \node (h) at (3,-0.3) {#2};
\end{tikzpicture}
}

\newcommand{\scaleNamedInfernoWidth}[4]
{
\def\colorscale{figures/colorscale_inferno.png}
\begin{tikzpicture}
      \node[anchor=south west,inner sep=0] at (0,0)
      {\includegraphics[width={#4}, height=0.3cm]{\colorscale}};
      \node (h) at (0,-0.3) {#1};
      \node (h) at ({#4/2},-0.3){#3};
      \node (h) at ({#4},-0.3) {#2};
\end{tikzpicture}
}

\newcommand{\scaleNamedInfernoHeight}[4]
{
\def\colorscale{figures/colorscale_inferno.png}
\begin{tikzpicture}
      \node[anchor=south west,inner sep=0] at (0,0)
      {\includegraphics[width={#4}, height=0.3cm,angle=90,origin=c]{\colorscale}};
      \node[anchor=west] (h) at (0.3,0) {#1};
      \node[anchor=west] (h) at (0.3,{#4/2}){#3};
      \node[anchor=west] (h) at (0.3,{#4}) {#2};
\end{tikzpicture}
}

\newcommand{\reffig}[1]{Fig.~\ref{#1}}
\newcommand{\refeqn}[1]{Equation~(\ref{#1})}
\newcommand{\reftab}[1]{Table~\ref{#1}}

\newcommand{\refsec}[1]{Section~\ref{#1}}

\setkeys{glslink}{hyper=false}

\renewcommand*{\CustomAcronymFields}{%
  name={\the\glsshorttok},%
  description={\the\glslongtok},%
}

\SetCustomStyle
\makeatletter
\def\ps@pprintTitle{%
 \let\@oddhead\@empty
 \let\@evenhead\@empty
 \def\@oddfoot{}%
 \let\@evenfoot\@oddfoot}
\makeatother

\begin{document}

\begin{frontmatter}


\title{Standardized Non-Intrusive Reduced Order Modeling Using Different Regression Models With Application to Complex Flow Problems}

\author[add1]{Art\={u}rs B\={e}rzi\c{n}\v{s}\corref{cor1}}
\ead{berzins@cats.rwth-aachen.de}
\author[add1]{Jan Helmig\corref{cor1}}
\ead{helmig@cats.rwth-aachen.de}
\author[add1,add2]{Fabian Key}
\ead{key@ilsb.tuwien.ac.at}
\author[add1,add2]{Stefanie Elgeti}
\ead{elgeti@ilsb.tuwien.ac.at}

\cortext[cor1]{Corresponding authors}

\address[add1]{Chair for Computational Analysis of Technical Systems (CATS)
    \\ RWTH Aachen University, 52056 Aachen, Germany}
\address[add2]{Institute of Lightweight Design and Structural Biomechanics
    \\ TU Wien, A-1060 Vienna, Austria}

\begin{abstract}
In recent years, numerical methods in industrial applications have evolved from a pure predictive tool towards a means for optimization and control. Since standard numerical analysis methods have become prohibitively costly in such multi-query settings, a variety of \gls{ROM} approaches have been advanced towards complex applications. In this context, the driving application for this work is \glspl{TSE}: manufacturing devices with an important economic role in plastics processing. Modeling the flow through a \gls{TSE} requires non-linear material models and coupling with the heat equation alongside intricate mesh deformations, which is a comparatively complex scenario. We investigate how a non-intrusive, data-driven \gls{ROM} can be constructed for this application. We focus on the well-established \gls{POD} with regression albeit we introduce two adaptations: standardizing both the data and the error measures as well as – inspired by our space-time simulations – treating time as a discrete coordinate rather than a continuous parameter. We show that these steps make the \gls{POD}-regression framework more interpretable, computationally efficient, and problem-independent. We proceed to compare the performance of three different regression models: \Gls{RBF} regression, \gls{GPR}, and \glspl{ANN}. We find that \gls{GPR} offers several advantages over an \gls{ANN}, constituting a viable and computationally inexpensive non-intrusive \gls{ROM}. Additionally, the framework is open-sourced \footnote{\url{https://github.com/arturs-berzins/sniROM}} to serve as a starting point for other practitioners and facilitate the use of \gls{ROM} in general engineering workflows.
\end{abstract}



\end{frontmatter}

\section{Introduction}
\label{sec:Introduction}

Many problems in engineering and science are modeled as \glspl{PDE} that may be parametrized in material properties, initial and boundary conditions or geometry.
Numerical methods, such as the \gls{FEM}, have been widely adopted to solve these problems.
However, obtaining a high-fidelity solution for complex problems is demanding in terms of the required computational resources. Especially in many-query contexts, such as optimization or uncertainty-quantification, where the \gls{PDE} is repeatedly solved for different parameters, the computational burden becomes impractical. Similarly, in time-critical applications, e.g. control, the real-time evaluation of a complex model requires prohibitive amounts of computational power and storage.

\Gls{ROM} is an umbrella term for methods aiming to alleviate this computational cost by replacing the full-order system by one with a significantly smaller dimension and paying a price of a controlled loss in accuracy \cite{Chen2018}.

\Gls{ROM} methods have been developed simultaneously in different fields of research, notably control, structural mechanics and fluid dynamics \cite{MOR2015, Rozza2018, Benner2015}. As a result, a multitude of \gls{ROM} methods and classifications thereof have developed.
Antoulas et al. \cite{MOR2015} classify \gls{ROM} in 
\emph{truncation} based methods, which focus on preserving key characteristics of the system rather than reproducing the solution, and
\emph{projection} based methods, which replace the high dimensional solution space with a reduced space of a much smaller dimension. 
Eldred and Dunlavy \cite{Eldred2006} and Benner et al. \cite{Benner2015} use three categories instead: \emph{hierarchical}, \emph{data-fit}, and \emph{projection-based} reduced models. The former category includes a range of physics-based approaches, although other authors exclude simplified-physics methods from \gls{ROM} altogether \cite{Rozza2018, malik2017}. 
Data-fit models use interpolation or regression methods to map system inputs to outputs, i.e., parameters to quantities of interest. 
Finally, according to this classification, in projection-based methods, the full-order operators are projected onto a reduced basis space allowing to solve a small reduced model quickly.
 
 In other communities, including \gls{CFD}, \gls{ROM} is typically divided into \emph{intrusive} and \emph{non-intrusive} methods \cite{Yu2019, Hesthaven2018, Guo2018, Guo2019, Rozza2018}.
 Intrusive \gls{ROM} corresponds to the previous definition of projection based methods, while in non-intrusive \gls{ROM} the solutions instead of the operators are projected onto the reduced basis. This yields a compact representation of a solution in the reduced basis as a vector of \emph{reduced coefficients}.
 Regression or interpolation can then be applied to rapidly determine the reduced coefficients at a new parameter instance. This makes the non-intrusive \gls{ROM} fall into the previously defined \emph{data-fit} class.
 
 Both intrusive and non-intrusive \gls{ROM} methods are characterized by an offline-online paradigm. The \emph{offline} phase is associated with some computational investment due to generating a collection of solutions, extracting the reduced basis and setting up the \gls{ROM}. However, this allows to rapidly evaluate the \gls{ROM} during the \emph{online} stage. Ideally, the complexity of the online evaluation is independent of the full-order model.
 
 Both \gls{ROM} approaches also share the question of how to determine the reduced basis.
 A multitude of methods have been proposed in the literature such as greedy algorithms, dynamic mode decomposition, autoencoders and others \cite{DMDbook2016, Yu2019, Benner2015}. However, the arguably most popular method is \gls{POD}, which constructs a set of orthonormal basis vectors representing common modes in a collection of solutions \cite{Yu2019}. 

Within intrusive \gls{ROM}, mostly the \emph{Galerkin} procedure is used \cite{Yu2019, HesthavenBook2015, Hesthaven2018, Guo2018}. However, a naive approach of projecting the operators onto the reduced basis space has a crucial flaw for non-linear problems, namely, the parameter dependent full-order operators still have to be reassembled during the \emph{online} computation. This severely limits rapid evaluation of new reduced solutions.
\emph{Affine expansion} mitigates this problem under the assumption of \emph{affine decomposability} of the operators in the weak form. However, this assumption is violated for general \emph{non-affine} problems \cite{HesthavenBook2015}.
Approaches like the \gls{EIM} \cite{Barrault2004}, discrete \gls{EIM} \cite{DEIM2010} and \gls{TPWL} method \cite{Rewieski2006} have been introduced to recover the advantage of an affine decomposition by another approximation \cite{Yu2019, Benner2015}, but they are problem-dependent and often impractical for general non-linear problems \cite{Guo2018}.

An alternative approach is offered by the \emph{non-intrusive} methods. These enable a purely \emph{data-driven} approach, as the solution set for the offline phase can originate from an unmodified solver or even experimental data.
By projecting a solution onto the reduced basis, typically acquired via \gls{POD}, the solution is compactly represented as a vector of reduced coefficients.
The key step in the non-intrusive framework is fitting a regression model that maps the parameters to reduced coefficients.
In principle, any interpolation or regression method can be used with \gls{POD}, such as,
least squares regression \cite{Dolci2016} or
cubic spline interpolation \cite{BuiThanh2003ProperOD},
but more common methods in the literature are
\gls{RBF} interpolation \cite{Rozza2018, Dolci2016, Walton2013ReducedOM, Xiao2015, Iuliano2013, Xiao2017},
\gls{GPR} \cite{Guo2018, Guo2019, Dupuis2018} and
recently \glspl{ANN} \cite{Hesthaven2018, Wang2019, Gao2019, Park2013}.

In the \gls{CFD} context, non-intrusive \glspl{ROM} have been applied to a variety of problems.
Examples of \gls{POD}-\gls{GPR} applications include time-dependent one-dimensional Burgers' equation \cite{Guo2019}, incompressible fluid flow around a cylinder \cite{Guo2019}, and moving shock in a transonic turbulent flow \cite{Dupuis2018}.
\gls{POD}-\gls{ANN} has been used for
quasi-one dimensional unsteady flows in continuously variable resonance combustors \cite{Wang2019}, 
steady incompressible lid-driven skewed cavity problem \cite{Hesthaven2018}, 
convection dominated flows with application to Rayleigh-Taylor instability \cite{Gao2019},
transient flows described by one-dimensional Burgers and two-dimensional Boussinesq equations \cite{Omer2018} and 
aerostructural optimization \cite{Park2013}. 
Numerous examples of \gls{POD}-\gls{RBF} also exist \cite{Rozza2018, Dolci2016, Walton2013ReducedOM, Xiao2015, Iuliano2013}.
For a more complete overview on \gls{ROM} in \gls{CFD} we refer to existing reviews \cite{Yu2019, Rozza2018}.

All examples named above are restricted to Newtonian fluids. In contrast, only a few works apply \gls{ROM} to \emph{generalized Newtonian fluids}. These include the \gls{TPWL} method for transient elastohydrodynamic contact problems \cite{Maier2015}, 
\gls{POD}-Galerkin method for a steady incompressible flow of a pseudo-plastic fluid in a circular runner \cite{Girault2018}
and \gls{ROM} based on residual minimization for generic power-law fluids \cite{Ocana2014}.
All these works use intrusive methods. In \cite{muravleva2018application}, a non-intrusive \gls{ROM} based on \gls{ANN} has been applied to viscoplastic flow modeling.

In this work, we construct a non-intrusive \gls{ROM} framework using \gls{POD} with regression and apply it to three different complex flow problems.
Three different regression models, namely, \gls{RBF} regression, \gls{GPR} and \glspl{ANN}, are employed and compared.
We emphasize several adjustments to the currently established \gls{POD}-regression framework:
\begin{enumerate}
    \item centering before \gls{POD},
    \item standardizing by singular values before regression, and
    \item the use of a standardized error measure.
\end{enumerate}
These steps make the \gls{POD}-regression framework more interpretable by establishing connections to known statistical concepts as well as more problem-independent and computationally efficient due to the regression maps having a more predictable behaviour.

Additionally, for time-resolved problems, in contrast to the established approach of treating time as a continuous parameter \cite{Yu2019, Audouze2013, Chen2018, Xiao2015, Guo2019, Wang2019, Peherstorfer2016}, we propose to treat time as a discrete coordinate.
Our approach allows us to address both steady and unsteady problems with the same framework while also preserving computational efficiency of both \gls{POD} and regression.

Our method is first validated on both steady and time-dependent lid-driven cavity viscous flows. 
Finally, the same framework is applied to a cross-section of a co-rotating twin-screw extruder, which is characterized by a time- and temperature-dependent flow of a generalized Newtonian fluid on a deforming domain.
Twin-screw extruders are important devices used in the plastics industry for performing multiple processing operations simultaneously, including melting, compounding, blending, pressurization and shaping. Their importance stems from the fact that they provide a short residence time, extensive mixing, and a modular design adaptable to individualized polymers. To perform process optimization, a good understanding of the temperature distributions, mixing behaviours and residence times inside the extruder are necessary. However, experimental investigation is difficult due to the complex moving parts, small gap sizes and high pressures \cite{Helmig2019}.
\gls{CFD} offers an appealing alternative, however, the computational burden in a many-query evaluation of the extruder is very high. As a consequence, we investigate the potentials offered by \gls{ROM}.

This work is structured as follows: \refsec{sec:Non-intrusive reduced-basis method} formally introduces the non-intrusive \gls{ROM}, the \gls{POD}-regression framework, and the proposed adjustments.
\refsec{sec:Regression models} describes the used \gls{RBF}, \gls{GPR} and \gls{ANN} regression models.
\refsec{sec:NS&solver} outlines the governing equations and numerical method used to compute the datasets.
In \refsec{sec:results} our non-intrusive \gls{ROM} is first validated against existing results on the skewed lid-driven cavity problem. Afterwards, the framework is transferred to the oscillating lid-driven cavity problem and, finally, the twin-screw extruder.
\section{Non-intrusive reduced-basis method}
\label{sec:Non-intrusive reduced-basis method}

After providing a high-level overview on \gls{ROM} in \refsec{sec:Introduction}, we describe the non-intrusive \gls{ROM} using \gls{POD} and regression more formally in the following.
We first describe the method as it is commonly reported in the literature \cite{Hesthaven2018, Guo2018, HesthavenBook2015, Yu2019}. Then in Sections \ref{sec:preprocessing}-\ref{sec:unsteadyRB}, we detail our proposed modifications.

Although the data for the non-intrusive \gls{ROM} can originate from any numerical scheme or even experimental measurements, we illustrate the purpose of a \emph{basis} using \gls{FEM}. The high-fidelity solution to a parametrized \gls{PDE} provided by an \gls{FEM} solver is typically of the form 
\begin{equation}
 \label{eq:FEM}
  s (\vekt{x}; \vekt{\mu}) 
  = \vekt{s}^\top(\vekt{\mu}) \vekt{\phi}(\vekt{x})
  = \sum_{i=1}^{N_h} s_i (\vekt{\mu})  \phi_i(\vekt{x}) \ , 
\end{equation}

where
$ \vekt{\phi} = [ \phi_{1}(\vekt{x}) | \cdots | \phi_{N_h}(\vekt{x}) ]^\top \in \mathbb{R}^{N_h}$
is a collection of the (e.g. Lagrange) basis functions and the solution vector $\vekt{s} (\vekt{\mu}) \in \mathbb{R}^{N_h}$ fixes the coefficient values for all $N_h$ \glspl{DOF}.
The vector $\vekt{\mu} \in \mathbb{R}^{N_d}$ collects all $N_d$ input parameters.
Given a training set with $N_{tr}$ different parameter samples $\mathbb{P}_{tr} = \{ \vekt{\mu}^{(n)} \}_{1 \leq n \leq N_{tr}}$
and corresponding high-fidelity solution coefficients
$\mathbb{S}_{tr} = \{ \vekt{s}(\vekt{\mu}) \}_{\vekt{\mu} \in \mathbb{P}_{tr}}$,
the key idea in \gls{ROM} is to construct a set of $L$ \emph{reduced basis} vectors 
$\matr{V} = [ \vekt{v}^{(1)} | ... | \vekt{v}^{(L)}] \in \mathbb{R}^{N_h \times L}$
such that for any solution vector $\vekt{s} (\vekt{\mu})$ an approximation $\vekt{s}_L (\vekt{\mu})$ can be constructed as a linear combination of a small number of reduced basis vectors $L \ll N_h$: 
  \begin{equation}
  \label{eq:approx}
    \vekt{s} (\vekt{\mu})
    \approx {\vekt{s}}_L (\vekt{\mu})
    = \sum_{l=1}^{L} y_{l} (\vekt{\mu}) \vekt{v}^{(l)}
    = \matr{V}\vekt{y}(\vekt{\mu}) \ ,
  \end{equation}

with the corresponding \emph{reduced coefficients} $\vekt{y} (\vekt{\mu}) =  [y_l|\cdots|y_L]^\top \in \mathbb{R}^L$.
Inserting \refeqn{eq:approx} in \refeqn{eq:FEM} gives the spatially interpolated approximation:
\begin{equation}
    s(\vekt{x};\vekt{\mu}) \approx s_L(\vekt{x};\vekt{\mu}) = \vekt{y}^\top(\vekt{\mu}) \matr{V}^\top \vekt{\phi}(\vekt{x}) \ .
\end{equation}
We refer to $\matr{V}^\top \vekt{\phi}(\vekt{x})$ as the reduced basis \emph{functions}. However, throughout this work, we mostly operate on the discrete entities $\matr{V}$ and $\vekt{s}(\vekt{\mu})$ and refer to them simply as \emph{reduced basis} and \emph{solution}, respectively.

A commonly used tool to compute the reduced basis is \gls{POD} (see \refsec{sec:POD}), which generates a set of orthonormal reduced basis vectors such that $\matr{V}^\top\matr{V}=\matr{I}$. As a result, the reduced coefficients can be computed by \emph{projecting} the high-fidelity solutions onto the reduced basis: 
\begin{equation}
 \label{eq:rbcoeffs}
 \vekt{y} (\vekt{\mu}) = \matr{V}^\top\vekt{s}(\vekt{\mu}) \ .
\end{equation}

The final step of the offline stage is to recover the unknown underlying mapping from parameters to the reduced coefficients $\vekt{\pi}: \vekt{\mu} \mapsto \vekt{y} (\vekt{\mu}) $. 
This is done by fitting a regression model $\tilde{\vekt{\pi}}$ on the training data consisting of the input set of parameter samples $\mathbb{P}_{tr}$ and the output set of the corresponding reduced coefficients $\mathbb{Y}_{tr}=\{\vekt{y}(\vekt{\mu})\}_{\vekt{\mu} \in \mathbb{P}_{tr}}$.
 Several choices for regression models are described in Section \ref{sec:Regression models}. During the online stage, the fitted regression model can be evaluated rapidly for new parameter samples to obtain the \emph{predicted reduced coefficients} $\tilde{\vekt{y}}$. These can be transformed back to the full space to reconstruct the \emph{predicted solution} $\tilde{\vekt{s}}_L$:
 
 \begin{equation}
\label{eq:up-projection}
 \tilde{\vekt{s}}_L (\vekt{\mu}) = \matr{V} \tilde{\vekt{y}}(\vekt{\mu}) = \matr{V} \tilde{\vekt{\pi}}(\vekt{\mu}) \ .
\end{equation}

The recovery requires $\mathcal{O}(N_h L)$ operations due to the matrix-vector multiplication, which involves the number of \glspl{DOF} in the full-order problem $N_h$.
Depending on the application, this may, however, not even be required either if the quantity of interest is computed directly from the reduced coefficients or if subsampling of the solutions can be used.
Note that the online evaluation is completely independent of the solver used during the offline phase. 
In the following, we introduce the concepts of \gls{POD} and standardization.

\subsection{Proper orthogonal decomposition}
\label{sec:POD}

Given a \emph{snapshot matrix} $ \matr{S} = [ \vekt{s} (\vekt{\mu}^{(1)}) | ... | \vekt{s} (\vekt{\mu}^{(N_{tr})}) ] \in \mathbb{R}^{N_h \times N_{tr}}$ in which the high-fidelity solutions are arranged column-wise,
\gls{POD} makes use of the \gls{SVD} to decompose the normalized snapshot matrix $\matr{S}/\sqrt{N_{tr}}$ into two orthonormal matrices $\matr{W} \in \mathbb{R}^{N_h \times N_h}$, $\matr{Z} \in \mathbb{R}^{N_{tr} \times N_{tr}}$ and a diagonal matrix $\matr{\Sigma} \in \mathbb{R}^{N_h \times N_{tr}}$ such that

\begin{equation}
 \matr{S}/\sqrt{N_{tr}} = \matr{W} \matr{\Sigma} \matr{Z}^\top \ .
\end{equation}

 Columns of $\matr{W} = [\vekt{w}_1 | \cdots | \vekt{w}_{N_h}]$ and $\matr{Z} =  [\vekt{z}_1 | \cdots | \vekt{z}_{N_{tr}}]$ are the \emph{left} and \emph{right singular vectors} of both $\matr{S}$ and $\matr{S}/\sqrt{N_{tr}}$. 
The entries in the rectangular diagonal matrix $\matr{\Sigma}=diag({\sigma_1, \cdots, \sigma_N})$ are the \emph{singular values} of $\matr{S}/\sqrt{N_{tr}}$ and are ordered decreasingly $\sigma_1 \ge \sigma_2 \ge \cdots \ge \sigma_{N} \ge 0 $.
The \emph{closeness} between a snapshot $\vekt{s} (\vekt{\mu})$ and its rank $L$ approximation $ \vekt{s}_L (\vekt{\mu}) $ in the basis $\matr{V}$ can be quantified as their Euclidian distance using Equations (\ref{eq:approx}) and (\ref{eq:rbcoeffs}):

\begin{equation}
\label{eq:delta}
\delta_\text{POD}(\vekt{\mu}; \matr{V}) = 
|| \vekt{s}(\vekt{\mu}) - \vekt{s}_L(\vekt{\mu}) ||_2 = 
|| \vekt{s}(\vekt{\mu}) - \matr{V}\matr{V}^\top \vekt{s}(\vekt{\mu}) ||_2 \ .
\end{equation}

The \gls{SEY} \cite{Schmidt1907, Eckart1936} theorem states that the first $L$ left singular vectors of $\matr{S}$, i.e., $[\vekt{w}_1 | \cdots | \vekt{w}_{L}]$ are the optimal choice among all orthonormal bases $\hat{\matr{V}} \in \mathbb{R}^{N_h \times L}$

\begin{equation}
\label{eq:delta_train}
 \matr{V} = [\vekt{w}_1 | \cdots | \vekt{w}_L] = \operatorname*{argmin}_{ \hat{\matr{V}} } \delta_\text{POD}(\mathbb{P}_{tr};\hat{\matr{V}})
\end{equation}

with respect to minimizing the root-mean-square of the projection error over all training snapshots
\begin{equation}
\label{eq:Frobenius}
\delta_\text{POD}(\mathbb{P}_{tr}; \hat{\matr{V}}) 
= \sqrt{\frac{1}{N_{tr}} \sum_{\vekt{\mu} \in \mathbb{P}_{tr}} \delta^2_\text{POD}( \vekt{\mu}; \hat{\matr{V}} ) } 
= \frac{1}{\sqrt{N_{tr}}}
|| \matr{S} - \hat{\matr{V}}\hat{\matr{V}}^\top \matr{S} ||_F \ ,
\end{equation}

where $||\cdot||_F$ denotes the Frobenius matrix norm.
The notation $\delta_\text{POD}(\mathbb{P}_{tr})$ is a generalization of $\delta_\text{POD}(\{\vekt{\mu}\})$ = $\delta_\text{POD}(\vekt{\mu})$, where we simply drop the brackets of the singleton set for ease of notation.
The \gls{SEY} theorem also states that the minimal associated projection error can be expressed as the root-squared sum of the left out singular values 
\begin{equation}
\label{eq:delta_sumsigma}
  \delta_\text{POD}(\mathbb{P}_{tr} ;\matr{V}) = \sqrt{\sum_{l=L+1}^{N_{tr}} \sigma_l^2} \ .
\end{equation}

Computing the \gls{SVD} directly is prohibitively expensive, so the more efficient \emph{method of snapshots} is used in practice and in this work \cite{Sirovich}.
The idea is to first compute the eigenvalue decomposition of either $\matr{S}\matr{S}^\top /N_{tr} \in \mathbb{R}^{N_h \times N_h}$ or $ \matr{S}^\top\matr{S} /N_{tr} \in \mathbb{R}^{N_{tr} \times N_{tr}} $, depending on which one is smaller. This limits the computational complexity of \gls{POD} to be at worse cubic in the minimum of these dimensions $\mathcal{O}(\min\left\{N_{tr}, N_h\right\}^3)$ \cite{Pan1999}.
 Typically, the number of nodes $N_h$ is much larger than number of snapshots $N_{tr}$, so the eigendecomposition works out as $\matr{S}^\top\matr{S}/N_{tr} = \matr{Z}\matr{\Lambda}\matr{Z}^\top$, with $\matr{Z} \in \mathbb{R}^{N_{tr} \times N_{tr}}$ collecting the orthonormal eigenvectors and $\matr{\Lambda} = diag(\lambda_1, \cdots, \lambda_N)$ containing the real and positive eigenvalues of $\matr{S}^\top\matr{S}/N_{tr}$. With the above definition of \gls{SVD}, it follows that $\matr{\Sigma} = \matr{\Lambda}^{1/2}$ and $\matr{W} = \matr{S}\matr{Z}\matr{\Sigma}^{-1} / \sqrt{N_{tr}}$.

\subsection{Standardization}
\label{sec:preprocessing}
In the following, we describe the proposed \emph{standardization} steps to the non-intrusive \gls{ROM}. We start by providing theoretical bounds and motivation for the effectiveness of snapshot centering. Next, we show how standardization by singular values -- considered one of 'Tricks of the Trade' in machine learning \cite{LeCun} -- is a natural extension to the \gls{POD}-regression framework, which makes full use of the \gls{POD} and relieves the regression process.

\subsubsection{Snapshot centering}
\label{sec:centering}
Within the \gls{ROM} community, there is no clear consensus on whether to perform \gls{POD} on \emph{non-centered} data as described in \refsec{sec:POD} or on \emph{centered} data. The \emph{centered snapshot matrix}
$\matr{S}^c = [ \vekt{s}^c (\vekt{\mu}^{(1)}) | \cdots | \vekt{s}^c (\vekt{\mu}^{(N_{tr})}) ]$
is obtained by subtracting the mean over all columns $\bar{\vekt{s}} = 1/N_{tr} \sum_{\vekt{\mu} \in \mathbb{P}_{tr}} \vekt{s}(\vekt{\mu})$ from each snapshot:
\begin{equation}
\label{eq:q_h^c}
 \vekt{s}^c (\vekt{\mu}) = \vekt{s} (\vekt{\mu}) - \bar{\vekt{s}} \ .
\end{equation}

Some authors suggest that centering before \gls{POD} is 'customary' in \gls{ROM} \cite{Xiao2015, Ahmed2020, DMDbook2016}, but many counterexamples to this can also be found \cite{Guo2018, Hesthaven2018, Wang2019, Yu2019}. 
In \cite{Ahmed2020} it is even argued for and in \cite{Chen2012} against centering based on empirical evidence from specific applications.
We prefer centering based on the following motivation.

According to the \gls{SEY} theorem, the optimal reduced basis for centered data $\matr{V}^c$ still consists of the left-hand singular vectors of $\matr{S}^c$ and the committed projection error can still be expressed as the sum of the truncated modes:

\begin{equation}
\label{eqn:delta_POD_centered}
\delta_\text{POD}(\mathbb{P}_{tr}; {\matr{V}^c}) =
 \sqrt{
  1/N_{tr}
  \sum_{\vekt{\mu} \in \mathbb{P}_{tr}} ||
  \vekt{s}^c(\vekt{\mu}) -
  \matr{V}^c {\matr{V}^c}^\top
  \vekt{s}^c(\vekt{\mu})
  ||_2^2  
  }
  =
  \sqrt{\sum_{l=L+1}^{N_{tr}} (\sigma^c_l)^2}
   \ .
\end{equation}

After centering, ${\matr{S}^c}^\top\matr{S}^c/N_{tr}$ becomes the \emph{covariance} matrix and the singular values $\sigma^c$ of $\matr{S}^c/\sqrt{N_{tr}}$ can readily be interpreted as the population standard deviations along the respective reduced basis vectors.

Honeine \cite{Honeine2014} discusses the effects of applying \gls{SVD} to centered data. Importantly, Theorem 3 in \cite{Honeine2014} establishes upper and lower bounds for singular values $ \sigma_{l+1} \le \sigma^c_{l} \le \sigma_{l}$ for all $1 \le l \le N_{tr}$ from which bounds for the projection errors follow:
\begin{equation}
\delta_\text{POD}(\mathbb{P}_{tr}; \matr{V}_{L+1}) \le
\delta_\text{POD}(\mathbb{P}_{tr}; \matr{V}^c_L) \le
\delta_\text{POD}(\mathbb{P}_{tr}; \matr{V}_L).
\end{equation}

This means that with respect to the projection error, centering can effectively save one basis function, but also no more than that. According to Theorem 1 in \cite{Honeine2014}, the difference is diminishing for large $L$, but significant for small $L$:
\begin{equation}
 \delta^2_\text{POD}(\mathbb{P}_{tr}; \matr{V}_{L}) -
\delta^2_\text{POD}(\mathbb{P}_{tr}; \matr{V}^c_{L}) =
\begin{cases}
N_{tr}||\bar{\vekt{s}}||_2  & \text{for } L=0 \\
0                           & \text{for } L=N_{tr} \ .
\end{cases}
\end{equation}
 

Furthermore, centering also guarantees that the \emph{centered reduced coefficients} $\vekt{y}^c(\vekt{\mu})=\matr{V}^c \vekt{s}^c(\vekt{\mu})$ are also zero mean: $ 1/N_{tr} \sum_{\vekt{\mu} \in \mathbb{P}_{tr}} \vekt{y}^c(\vekt{\mu})  = \vekt{0} $, which is a common assumption in \gls{GPR} (see \refsec{sec:GPR}).

\subsubsection{Coefficient standardization}
\label{sec:scaling}

Another approach adopted in this work is coefficient standardization. The \emph{standardized reduced coefficients} are obtained by scaling the coefficients by the corresponding standard deviations:
\begin{equation}
    \vekt{y}^{s}(\vekt{\mu}) = (\matr{\Sigma}^c)^{-1} \vekt{y}^{c} (\vekt{\mu}) \ .
\end{equation}

\begin{figure}[tbp]
    \centering
    \tikzset{>=latex}
\begin{tikzpicture}[
    font=\footnotesize,
	scale=1,
	declare function={
		samplenormal(\u,\v)=sqrt(-2*ln(\u))*sin(2*pi*\v r);
	}]

\def\A{2}			
\def\B{2.4}			
\def\D{(\A+\B)}		
\def\whereX{below}	
\def\whereY{left}	
\def\pad{.6}		

\def\sigx{1.5}	
\def\sigy{.5}	
\def\a{30}		
\def\cx{0.75}		
\def\cy{1.5}		

\def\s{3} 
\def\k{1} 
	
\foreach\i in {0,...,99}{
	
	%
    	\pgfmathsetmacro{\x}{samplenormal(rnd, rnd)/\s}
    	\pgfmathsetmacro{\y}{samplenormal(rnd, rnd)/\s}
    	\fill [shift={({3*\D}, 0)}, red, opacity=0.1] (\x,\y) circle [radius=0.1];

	\pgfmathsetmacro{\xs}{\sigx*\x}
	\pgfmathsetmacro{\ys}{\sigy*\y}
	\fill [shift={({2*\D}, 0)}, red, opacity=0.1] (\xs,\ys) circle [radius=0.1];

	\pgfmathsetmacro{\xr}{cos(\a)*\xs-sin(\a)*\ys}
	\pgfmathsetmacro{\yr}{sin(\a)*\xs+cos(\a)*\ys}
	\fill [shift={({1*\D}, 0)}, red, opacity=0.1] (\xr,\yr) circle [radius=0.1];

	\pgfmathsetmacro{\xo}{\xr+\cx}
	\pgfmathsetmacro{\yo}{\yr+\cy}
	\fill [shift={({0*\D}, 0)}, red, opacity=0.1] (\xo,\yo) circle [radius=0.1];
	
};

\draw[shift={({0*\D}, 0)}, thin,->] (0,0) -- (\A,0)  node[\whereX] {$\vekt{e}_1$};
\draw[shift={({0*\D}, 0)}, thin,->] (0,0) -- (0,\A)  node[\whereY] {$\vekt{e}_2$};

\draw[shift={({0*\D}, 0)}, thin,->] (\A,1) -- +({\B-2*\pad},0) node[midway, above]{center};

\draw[shift={({1*\D}, 0)}, thin,->] (0,0) -- (\A,0)  node[\whereX] {$\vekt{e}_1$};
\draw[shift={({1*\D}, 0)}, thin,->] (0,0) -- (0,\A)  node[\whereY] {$\vekt{e}_2$};

\draw[shift={({1*\D}, 0)}, thin,->] (\A,1) -- +({\B-2*\pad},0) node[midway, above]{project};

\draw[thin,->, shift={({2*\D}, 0)}] (0,0) -- (\A,0)  node[\whereX] {$\vekt{e}_1$};
\draw[thin,->, shift={({2*\D}, 0)}] (0,0) -- (0,\A)  node[\whereY] {$\vekt{e}_2$};

\draw[shift={({2*\D}, 0)}, thin,->] (\A,1) -- +({\B-2*\pad},0) node[midway, above]{standardize};

\draw[thin,->, shift={({3*\D}, 0)}] (0,0) -- (\A,0)  node[\whereX] {$\vekt{e}_1$};
\draw[thin,->, shift={({3*\D}, 0)}] (0,0) -- (0,\A)  node[\whereY] {$\vekt{e}_2$};


\draw[shift={({0*\D}, 0)}, thin,->] (0,0) -- (\cx,\cy) node[midway, right]{$\bar{\vekt{s}}$};

\draw[shift={({1*\D}, 0)}, thin,->] (0,0) -- ({ cos(\a)},{sin(\a)}) node[pos=1, anchor=north, sloped]{$\vekt{v}^c_1$};
\draw[shift={({1*\D}, 0)}, thin,->] (0,0) -- ({-sin(\a)},{cos(\a)}) node[pos=1, rotate=90, anchor=east, sloped]{$\vekt{v}^c_2$};

\draw[shift={({2*\D}, 0)}, thin] (0,0) ellipse ({\k*\sigx/\s} and {\k*\sigy/\s});
\node[shift={({2*\D}, -1.5em)}]  		at ({\k*\sigx/\s},0) {$\sigma^c_1$};
\node[shift={({2*\D}, .5em)}, left]  	at (0,{\k*\sigy/\s}) {$\sigma^c_2$};

\draw[shift={({3*\D}, 0)}, thin] (0,0) ellipse ({\k*1/\s} and {\k*1/\s});
\node[shift={({3*\D}, -1.5em)}]  		at ({\k*1/\s},0) {$1$};
\node[shift={({3*\D}, .5em)}, left]  	at (0,{\k*1/\s}) {$1$};

\node[shift={({0*\D}, 0)}, align=center]	at ({\A/2},-4em) {$\vekt{s}(\vekt{\mu})$};
\node[shift={({1*\D}, 0)}, align=center]	at ({\A/2},-4em) {$\vekt{s}^c(\vekt{\mu}) = \vekt{s}(\vekt{\mu}) - \bar{\vekt{s}}$};
\node[shift={({2*\D}, 0)}, align=center]	at ({\A/2},-4em) {$\vekt{y}^{c}(\vekt{\mu}) = {\matr{V}^c}^\top\vekt{s}^{c}(\vekt{\mu})$};
\node[shift={({3*\D}, 0)}, align=center]	at ({\A/2},-4em) {$\vekt{y}^{s}(\vekt{\mu}) = (\matr{\Sigma}^c)^{-1}\vekt{y}^{c}(\vekt{\mu})$};

\end{tikzpicture}
    \caption[Preprocessing steps.]{
    The figure illustrates the effect of the data preprocessing steps described in Section 2.2.
    For some parameter instance $\vekt{\mu}$, $\vekt{s}(\vekt{\mu})$ denotes the two-dimensional ($N_h=2$) snapshot vector.
    Each element in the dataset is illustrated with a red dot, where $\vekt{e}_i$ is the $i$-th unit vector.
    Data is projected on all available bases ($L=N_h=2$). 
    The final standardized data has zero mean and unit standard deviation along each basis vector, i.e., principal direction. The shift and scale invariance serves to improve interpretability and model hyperparameter reusability not only for different principal directions, but also across problems.}
    \label{fig:preprocessing}
\end{figure}
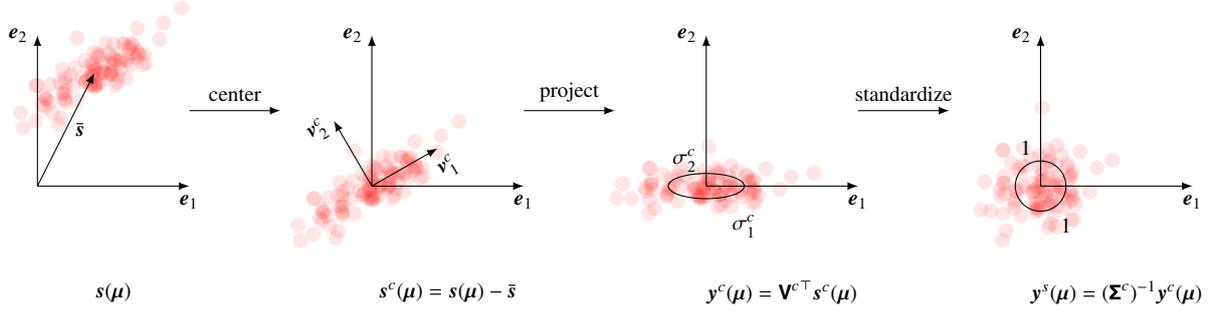

$\vekt{y}^{s}(\vekt{\mu})$ have a zero mean and unit variance for each of its $L$ components and can thus be viewed as a multi-variate Z-score of the centered reduced coefficients. 
This achieves dataset standardization across different problems and allows to reuse similar regression model architectures and learning processes, which are controlled by their \emph{hyperparameters}. Normally, these must be found using tedious trial-and-error or an extensive and computationally expensive hyperparameter-tuning. Standardization allows to set tight bounds on the search-space or even reuse hyperparameters (see \refsec{sec:Regression models}). 
\reffig{fig:preprocessing} helps illustrate the described transformation steps.

\subsubsection{Parameter standardization}
\label{sec:feature_standardization}

Similar to the standardized reduced coefficients, we also introduce the \emph{standardized parameters} as 
\begin{equation}
\vekt{\mu}^s = (\vekt{\mu} - \bar{\vekt{\mu}})/\bar{\bar{\vekt{\mu}}}\end{equation}
with the mean 
$\bar{\vekt{\mu}}
=
1/N_{tr} \sum_{\vekt{\mu} \in \mathbb{P}_{tr}} \vekt{\mu}
$
and the standard deviation 
$\bar{\bar{\vekt{\mu}}}
=
\sqrt{1/N_{tr} \sum_{\vekt{\mu} \in \mathbb{P}_{tr}} (\vekt{\mu} - \bar{\vekt{\mu}})^2}
$. These again have zero mean and unit variance on any dataset following the same distribution as the training datset.
This standardization step is beneficial for regression tasks in problems, where different parameters are on different scales (see e.g. \refsec{sec:extruder}). Both \gls{RBF} and \gls{GPR} rely on distances in parameter space and would otherwise neglect the smaller parameters (see \refsec{sec:RBF} and \refsec{sec:GPR}). Similarly, \glspl{ANN} show faster convergence with standardized inputs \cite{LeCun98}.

\subsection{Errors}
\label{sec:errors}

Within the \gls{POD}-regression framework, errors are used to describe the quality of the reduced basis and the regression map. In \refsec{sec:errors_absolute}, we show how a careful definition of these errors and their aggregation across multiple samples lead to an intuitive and efficiently computable total error.
In \refsec{sec:errors_standardized}, we propose to standardize these errors, again connecting the results to familiar statistical concepts, achieving better interpretability and comparability across datasets, as well as scale and translation invariances.

\subsubsection{Absolute errors}
\label{sec:errors_absolute}

The \emph{absolute projection error} $\delta_\text{POD}$ has been already introduced in 
\refeqn{eq:delta} as the Euclidian distance between the snapshot and its projection onto the reduced basis. The same holds with centering:
\begin{equation}
\delta_{\text{POD}}(\vekt{\mu}; \matr{V}^c) = 
|| \vekt{s}(\vekt{\mu}) - \vekt{s}_L(\vekt{\mu})||_2
=
|| \vekt{s}^c(\vekt{\mu}) - \matr{V}^c{\matr{V}^c}^\top \vekt{s}^c(\vekt{\mu}) ||_2 \ .
\end{equation}

Similarly, the \emph{absolute regression error} $\delta_\text{REG}$ between the prediction
$\tilde{\vekt{s}}_L(\vekt{\mu})$
and the projection
$\vekt{s}_L(\vekt{\mu})$ is introduced
to describe how close the learned regression map $\tilde{\vekt{\pi}}^s$ approximates the observed regression map $\vekt{\pi}^s : \vekt{\mu^s} \mapsto \vekt{y}^s$:

\begin{equation}
\label{eq:delta_REG}
\begin{split}
\delta_{\text{REG}}(\vekt{\mu}; \matr{V}^c, \tilde{\vekt{\pi}}^s)
={}&
|| \tilde{\vekt{s}}_L(\vekt{\mu}) - \vekt{s}_L(\vekt{\mu}) ||_2 
= 
|| \tilde{\vekt{s}}^c_L(\vekt{\mu}) - \vekt{s}_L^c(\vekt{\mu}) ||_2 
=
\\
={}&
|| \matr{V}^c\tilde{\vekt{y}}^c(\vekt{\mu}) - \matr{V}^c \vekt{y}^c(\vekt{\mu}) ||_2 
=
|| \tilde{\vekt{y}}^c(\vekt{\mu}) - \vekt{y}^c(\vekt{\mu}) ||_2
=
\\
={}&
|| \matr{\Sigma}^c \left(\tilde{\vekt{y}}^s(\vekt{\mu}) - \vekt{y}^s(\vekt{\mu})\right) ||_2
=
|| \matr{\Sigma}^c \left(\tilde{\vekt{\pi}}^s(\vekt{\mu}^s) - \vekt{\pi}^s(\vekt{\mu}^s)\right) ||_2
 \ .
\end{split}
\end{equation}

Lastly, the \emph{absolute total error} $\delta_\text{POD-REG}$ is introduced as the distance between the true and predicted solutions to quantify the performance of the whole non-intrusive \gls{ROM}:

\begin{equation}
\label{eq:delta_PODREG}
\delta_{\text{POD-REG}}(\vekt{\mu}; \matr{V}^c, \tilde{\vekt{\pi}}^s)
=
|| \vekt{s}(\vekt{\mu}) - \tilde{\vekt{s}}_L(\vekt{\mu}) ||_2
=
|| \vekt{s}^c(\vekt{\mu}) - \tilde{\vekt{s}}^c_L(\vekt{\mu}) ||_2 \ .
\end{equation}

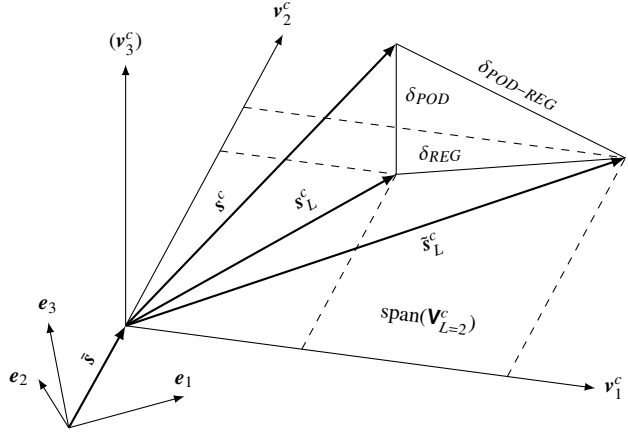
\begin{figure}[tbp]
\centering
\tikzset{>=latex}

\def \Tx {0.3} 
\def \Ty {0.6}	
\def \Tz {0.25}	
\def \Px {0.65}	
\def \Py {0.75}	

\tdplotsetmaincoords{60}{15}
\begin{tikzpicture}[scale=8, tdplot_main_coords, font=\footnotesize]

\draw[thin,->] (0,0,0) -- (.8,0,0) node[anchor= west]{$\vekt{v}^c_1$};
\draw[thin,->] (0,0,0) -- (0,1,0)  node[anchor=south]{$\vekt{v}^c_2$};
\draw[thin,->] (0,0,0) -- (0,0,.5) node[anchor=south]{$(\vekt{v}^c_3)$};

\draw[thick,->] (0,0,0) --(\Tx,\Ty,\Tz) node[pos=.4,sloped,above]{$\vekt{s}^c$};
\draw[thick,->] (0,0,0) --(\Tx,\Ty,0) node[pos=0.7,sloped, above]{$\vekt{s}^c_L$};
\draw[dashed] (\Tx,0,0) -- (\Tx,\Ty,0) node[]{};
\draw[dashed] (0,\Ty,0) -- (\Tx,\Ty,0) node[]{};

\draw[thick,->] (0,0,0) -- (\Px,\Py,0) node[pos=.6,sloped,below]{$\tilde{\vekt{s}}^c_L$};
\draw[dashed] (\Px,0,0) -- (\Px,\Py,0) node[]{};
\draw[dashed] (0,\Py,0) -- (\Px,\Py,0) node[]{};

\draw[thin](\Tx,\Ty,0) -- (\Tx,\Ty,\Tz) node[pos=.6, right]			{$\delta_{{POD}}$};
\draw[thin](\Tx,\Ty,  0) -- (\Px,\Py,0) node[pos=.2, sloped, above]	{$\delta_{{REG}}$};
\draw[thin](\Tx,\Ty,\Tz) -- (\Px,\Py,0) node[pos=.5, sloped, above]	{$\delta_{{POD-REG}}$};

\coordinate (Shift) at (-.07,-.1,-.15);
\tdplotsetrotatedcoordsorigin{(Shift)}

\tdplotsetrotatedcoords{0}{-10}{30}
\draw[thin,tdplot_rotated_coords,->] (0,0,0) --
(.2,0,0) node[anchor=south]{$\vekt{e}_1$};
\draw[thin, tdplot_rotated_coords,->] (0,0,0) --
(0,.2,0) node[anchor=east]{$\vekt{e}_2$};
\draw[thin,tdplot_rotated_coords,->] (0,0,0) --
(0,0,.2) node[anchor=south]{$\vekt{e}_3$};
\draw[thick,->] (Shift) -- (0,0,0) node[pos=.6,sloped,above]{$\bar{\vekt{s}}$};

\draw[draw=none] (0,.15,0) -- ({(\Tx+\Px)/2},.15,0) node[pos=1, sloped]	{$\operatorname{span}(\matr{V}^c_{L=2})$};

\end{tikzpicture}
\caption[Relation between errors.]{The total error $\delta_\text{POD-REG}$ can be decomposed into two orthogonal components -- the projection error $\delta_\text{POD}$ perpendicular to the span of centered basis $V^c$ and the regression error $\delta_\text{REG}$ belonging to the span. Illustration uses three degrees of freedom $N_h=3$ and two reduced basis vectors $L=2$.}
\label{fig:errors_illustrated}
\end{figure}

\reffig{fig:errors_illustrated} illustrates these three types of errors. Additionally, it can be seen that the total error is composed of two orthogonal components, namely the regression and projection errors:

\begin{equation}
\label{eq:delta_pythagorean}
{\delta_{\text{POD-REG}}(\vekt{\mu})}^2 = 
{\delta_{\text{POD}}(\vekt{\mu})}^2 + 
{\delta_{\text{REG}}(\vekt{\mu})}^2 \ .
\end{equation}

This can also be shown formally using the definition of the Euclidian vector norm $||\vekt{a}||_2^2=\vekt{a}^\top\vekt{a}$ and the fact that the predictions already belong to the linear span of the reduced basis $\matr{V}^c {\matr{V}^c}^\top \tilde{\vekt{s}}^c_L = \tilde{\vekt{s}}^c_L$:

\begin{equation}
\begin{split}
|| {\vekt{s}^c} - {\matr{V}^c}{\matr{V}^c}^\top{\vekt{s}^c} ||_2^2 +& || {\matr{V}^c}{\matr{V}^c}^\top{\vekt{s}^c} - {\tilde{\vekt{s}^c}_L} ||_2^2 =
\\ = 
  {{\vekt{s}^c}^\top}{\vekt{s}^c}
- {{\vekt{s}^c}^\top}{\matr{V}^c}&{{\matr{V}^c}^\top}{\tilde{\vekt{s}}^c_L} 
+ {{\tilde{\vekt{s}}^c_L}{}^\top} {\tilde{\vekt{s}}^c_L} =
\\ = 
  {{\vekt{s}^c}^\top}{\vekt{s}^c}
- {{\vekt{s}^c}^\top}&{\tilde{\vekt{s}}^c_L} 
+ {{\tilde{\vekt{s}}^c_L}{}^\top} {\tilde{\vekt{s}}^c_L} =
\\ = 
|| \vekt{s}^c -& {\tilde{\vekt{s}}^c_L} ||_2^2
\end{split}
\end{equation}

Lastly, we note that the equivalence in \refeqn{eq:delta_REG} paired with with \refeqn{eq:delta_pythagorean} allows to compute both $\delta_{\text{REG}}$ and $\delta_{\text{POD-REG}}$ in $\mathcal{O}(L)$ operations given that $\delta_{\text{POD}}$ is regression independent and can be pre-computed once.
In contrast to the naive approach via computing $\tilde{\vekt{s}}^c_L=\vekt{V}^c\tilde{\vekt{s}}^c$ requiring $\mathcal{O}(N_h L)$ operations due to matrix-vector multiplication, the presented approach allows to efficiently use and monitor all errors during the training of the regression models. 

The corresponding \emph{aggregate} projection, regression and total errors $\delta\left(\mathbb{P}\right)_*, *\in\left\{\text{POD},\text{REG},\text{POD-REG}\right\}$ on any dataset $\mathbb{P}$ of cardinality $N$ are defined via the \gls{RMSE}:
\begin{equation}
    \delta_*\left( \mathbb{P} \right)
    :=
    \sqrt{
    1/N \textstyle\sum_{\vekt{\mu} \in \mathbb{P}}
    {\delta_*\left( \vekt{\mu} \right)}^2
    } \ .
\end{equation}
This choice is motivated by the form of the \gls{SEY} theorem and treats both physical and parameter dimensions equally.
Hence, \refeqn{eq:delta_pythagorean} holds for both individual and aggregate errors.

\subsubsection{Standardized errors}
\label{sec:errors_standardized}

The use of absolute errors makes it difficult to assess and interpret the performance of \gls{ROM}, especially across different datasets.
Thus we introduce the corresponding \emph{standardized projection, regression} and \emph{total errors} in the following way:
\begin{equation}
    \varepsilon_*\left(\mathbb{P}\right) := \frac
    { \delta_*\left(\mathbb{P}\right) }
    {\sqrt{
        1/N_{tr} \sum_{\vekt{\mu} \in \mathbb{P}_{tr}}
        || \vekt{s}\left(\vekt{\mu}\right) - \bar{\vekt{s}} ||^2
    }} \ .
\end{equation}

The choice of the denominator is motivated by its relation to the standard deviation of the training set. Hence, $\varepsilon$ can be interpreted as a multivariate extension of \emph{Z-score} and describes the ratio between the Euclidian distance of two vectors and the standard deviation of the particular dataset.
Always naively predicting the mean vector $ \bar{\vekt{s}} $ leads to 100\% aggregated total error $\varepsilon_{\text{POD-REG}}(\mathbb{P}_{tr})=1$.
Furthermore, the denominator can be related to \gls{POD}, since
$ 1/N_{tr} \sum_{\vekt{\mu} \in \mathbb{P}_{tr}} 
|| \vekt{s}\left(\vekt{\mu}\right) - \bar{\vekt{s}} ||_2^2 
= 1/N_{tr} \Tr({\matr{S}^c}^\top \matr{S}^c)
= \sum_{n=1}^{N_{tr}}{(\sigma^c_n)^2}
$ .
Consequentially, it is easy to relate the standardized projection error to \gls{SEY} and \refeqn{eq:delta_sumsigma}:
\begin{equation}
{\varepsilon^2_\text{POD}(\mathbb{P}_{tr}; \matr{V}^c_L)} = 
\frac
{\sum_{n=L+1}^{N_{tr}}{(\sigma^c_n)^2}}
{\sum_{n=1}^{N_{tr}}{(\sigma^c_n)^2}}  \ .
\end{equation}
This corresponds to a commonly used measure 
known as the \emph{relative information loss}, which can be used to guide the choice of number of bases $L$. 
Furthermore, $\varepsilon$ is invariant to scaling and translation of snapshots and thus independent of the choice of units. This alleviates the need for designing dimensionless or normalized problems in order to interpret the errors and altogether allows for a more data-driven approach to non-intrusive \gls{ROM}.
Finally, within the broader classification of existing error measures in machine learning, our definition of $\varepsilon$ can be viewed as a multivariate extension of the root-relative-squared error, option 2 as classified by Botchkarev \cite{Botchkarev2018}.


\subsection{Time-resolved problems}
\label{sec:unsteadyRB}

In general, authors within \gls{ROM} community approach time-resolved problems by treating time as a continuous parameter \cite{Yu2019}. Wang et al. \cite{Wang2019} point out two issues that arise using this approach. Firstly, for problems with many time steps $N_t$, the snapshot matrix becomes wide $\matr{S} \in \mathbb{R} ^ {N_h {\times} N_{tr} N_t}$ and the cost of the \gls{POD} $\mathcal{O}\left( \min \left\{ N_h , N_{tr} N_t \right\}^3 \right)$ becomes high even using the method of snapshots because the smallest dimension is increased significantly.
Secondly, the regression model must be able to predict the solution at arbitrary time and parameter values. Thus it needs to learn both time and parameter dynamics.
Several solutions to these problems have been proposed, mainly using two-level \gls{POD} \cite{Audouze2013, Chen2018, Xiao2015, Guo2019, Wang2019} or operator inference \cite{Peherstorfer2016}. 

We propose an alternative method by relaxing the second requirement of needing to predict solutions at any arbitrary time instance. Instead, we treat time in discrete levels, similar to physical space.
This is inspired by typical numerical schemes, which also provide the solution at discrete spatial and temporal points, with any intermediate values being interpolated.
This approach ensures that the snapshot matrix becomes taller not wider $\matr{S} \in \mathbb{R} ^ {N_h N_t {\times} N_{tr} }$, preserving the smaller dimension $N_{tr}$ and ensuring efficiency of the \gls{POD}.
Each resulting basis vector spans space and time, and as a consequence the regression model must only account for parameter dynamics. This enables much simpler models and significantly reduces the effort of the fitting process.



\section{Regression models}
\label{sec:Regression models}

Within the non-intrusive \gls{ROM} framework, the trained regression model is used to predict the reduced coefficients at a new parameter location using  $\tilde{\vekt{\pi}}^s : \vekt{\mu}^s \mapsto \tilde{\vekt{y}}^s$. Note, that we standardize both the inputs and the outputs (see \refsec{sec:preprocessing}).
As described in \refsec{sec:Introduction}, any regression method can in principle be used. However, \gls{RBF} regression is commonly used due to its simplicity, but lately the more flexible \gls{GPR} and \gls{ANN} methods have been adopted in the \gls{ROM} community \cite{Yu2019, Rozza2018}.
These three models are briefly described in the following.



\subsection{Radial basis function regression}
\label{sec:RBF}

\gls{RBF} regression is a widely used tool in multivariate scattered data approximation.
A radial function $\Phi : \mathbb{R}^{N_d} \mapsto \mathbb{R}$ is multivariate, but can be reduced to a univariate function in some norm $||\cdot||$ of its argument: 
$\Phi(\vekt{\mu}^s) = \phi(||\vekt{\mu}^s||) \ \forall \vekt{\mu}^s \in \mathbb{R}^{N_d}$. Typically, the Euclidean norm $||\cdot||_2$ is used. 
\Gls{RBF} is \emph{radial} in the sense that the norm of the argument can be interpreted as a radius from the origin or the \emph{center}, i.e., $ r = ||\vekt{\mu}^s||_2 $.
In this work the \emph{multiquadric} \gls{RBF} $\phi(r) = \sqrt{1+(r/r_0)^2}$ is used. The hyperparameter $r_0$ controlling the scale is chosen as the mean distance between the training points in parameter space.

In a regression task, multiple \emph{translated} radial functions $\{ \phi(||\vekt{\mu}^s-\vekt{c}^{(m)}||) \}_{1 \le m \le M}$ are used to form a basis. The typical choice is to use as many basis functions as there are data-samples $ M = N_{tr} $ and to use the data locations as \emph{centers} $\vekt{c}^{(n)} = (\vekt{\mu}^s)^{(n)}$.
The approximating regression function $\tilde{\pi}^s_l: \mathbb{R}^{N_d} \mapsto \mathbb{R}, \ \tilde{\pi}^s_l (\vekt{\mu}) = \tilde{y}^s_l$ for a single output component $l \in [1,L]$ is expressed as a linear combination of the \glspl{RBF}:

\begin{equation}
\tilde{\pi}^s_l(\vekt{\mu}^s; \vekt{w}_l) = \sum_{m=1}^{M} \vekt{w}_l^{(m)} \phi\left(||\vekt{\mu}^s-(\vekt{\mu}^s)^{(m)}||\right) \ .
\end{equation}

The \emph{weights} $\vekt{w}_l$ can be determined by enforcing the \emph{exact interpolation condition} on the training data $\tilde{\pi}^s_l ( (\vekt{\mu}^s)^{(n)} ) = (y^s_l)^{(n)} \ \forall n \in [1,N_{tr}]$. This leads to a linear system $ \matr{A} \vekt{w}_l = \vekt{y}^s_l $ with $ \matr{A} = ( \phi(||(\vekt{\mu}^s)^{(n)}-(\vekt{\mu}^s)^{(m)}||) ) _{1 \le m,n \le N_{tr}}$ being the \emph{coefficient matrix}.

Note, that $\tilde{\pi}^s_l$ is a scalar function. For $L$ number of outputs, a different and independent function is constructed for each output dimension.
However, solving multiple linear systems $ \matr{A} \vekt{w}_l = \vekt{y}^s_l \text{ for } 1 \le l \le L$ is efficient, since they share the same coefficient matrix $\matr{A}$, which must be inverted or, in practice, decomposed only once. All the scalar interpolation functions can finally be gathered as components in the vector function $\tilde{\vekt{\pi}}^s = [\tilde{\pi}^s_1|\cdots|\tilde{\pi}^s_L]^\top$.

For more details we refer to \cite{Schaback2007}. 
The \gls{RBF} implementation used in our work is based on the SciPy module for Python \cite{SciPy-NMeth}.


\subsection{Gaussian process regression}
\label{sec:GPR}

In \gls{GPR}, again a scalar regression function $\tilde{\pi}^s_l:\mathbb{R}^{N_d} \mapsto \mathbb{R}, \ \tilde{\pi}^s_l(\vekt{\mu}^s)=\tilde{y}^s_l$ is constructed. However, instead of assuming a particular parametric form of the function and determining the parameters from the data, a \emph{non-parametric} approach is pursued, in which the probability distribution over all functions consistent with the observed data is determined. This distribution can then be used to predict the expected output at some unseen input.

A central assumption in \gls{GPR} is that \emph{any finite set} of outputs, gathered in the vector $\vekt{y}^s_l$, 
follows a multivariate Gaussian distribution
$ \vekt{y}^s_l \sim \mathcal{N}(\vekt{m}, \matr{K}) $
with the \emph{mean vector} $\vekt{m}$ and \emph{covariance matrix} $\matr{K}$. This informs the notion that the regression function is itself an 'infinite-dimensional' or 'continuous' multivariate distribution or, formally, a \emph{\gls{GP}}:
\begin{equation}
  \tilde{\pi}^s_l(\vekt{\mu}^s) \sim \mathcal{GP}( m(\vekt{\mu}^s),  k({\vekt{\mu}^s, \vekt{\mu}^s_\star}) ) \ .
\end{equation}

Here,
$(\vekt{\mu}^s, \vekt{\mu}^s_\star)$
are all possible pairs in the input domain and
$ m(\vekt{\mu}^s) =  \mathbb{E}[\tilde{\pi}^s_l(\vekt{\mu}^s)] $
and
$ k({\vekt{\mu}^s, \vekt{\mu}^s_\star})
 =
\mathbb{E}[
(\tilde{\pi}^s_l(\vekt{\mu}^s) - m(\vekt{\mu}^s)) (\tilde{\pi}^s_l(\vekt{\mu}^s_\star) - m(\vekt{\mu}^s_\star))
] $
are the \emph{mean} and \emph{covariance functions}, respectively. 
The mean vector $\vekt{m}$
and the covariance matrix $\matr{K}$
are thus just particular realizations of these functions at the $N$ corresponding input samples 
$\matr{P} = [ (\vekt{\mu}^s)^{(1)} | \cdots | (\vekt{\mu}^s)^{(N)} ] \in \mathbb{R}^{N_d \times N}$:
\begin{subequations}
\begin{align}
\vekt{m} &= m(\matr{P}) := [m((\vekt{\mu}^s)^{(i)})]_{1 \le i \le N} \in \mathbb{R}^{N}
\\
\matr{K} &= k(\matr{P}, \matr{P}) := [ k((\vekt{\mu}^s)^{(i)}, (\vekt{\mu}^s)^{(j)}) ]_{1 \le i,j \le N} \in \mathbb{R}^{N \times N} \ .
\end{align}
\end{subequations}

The central issue in \gls{GPR} is to determine a \emph{prior} on these functions, represented by the best \emph{belief} on the function's behaviour (e.g. smoothness) before any evidence is taken into account.
The prior is updated using training data to form a \emph{posterior} on the mean and covariance functions.
For the mean function, we adopt a widely used assumption of zero-mean $ m(\vekt{\mu}) = 0 $ \cite{Kuss2006, Rasmussen2005, Yu2019}. This corresponds to the actual prior belief since data centering is used \cite{Kuss2006} as described in \refsec{sec:centering}.

Determining the covariance function requires making stronger assumptions. The most widely adopted assumption is \emph{stationarity} of the inputs, expressing that the covariance between outputs only depends on the Euclidian distance between them in the input space. 
A common example is the \emph{squared exponential} function:
\begin{equation}
\label{eq:GPR_iso}
 k(\vekt{\mu}^s, \vekt{\mu}^s_\star; \sigma_f, d) = \sigma_f^2 \exp \left( -\frac{1}{2d^2} || \vekt{\mu}^s - \vekt{\mu}^s_\star ||^2 \right)
\end{equation}

with $\sigma_f$ being the prior covariance describing the level of uncertainty for predictions far away from training data and $d$ being the correlation lengthscale.
In this work, the more general \emph{anisotropic} squared exponential kernel is used, which prescribes unique correlation length $d_i$ for each input dimension and assumes additional observational noise $\varepsilon$ for numerical stabilization \cite{Kuss2006, Yu2019}:

\begin{equation}
\label{eq:GPR_aniso}
 k(\vekt{\mu}^s, \vekt{\mu}^s_\star; \sigma_f, \vekt{d}, \varepsilon) = \sigma_f^2 \exp \left( -\sum_{i=1}^{N_d}\frac{\left(\mu^s_i - {\mu^s_\star}_i\right)^2}{2d_i^2}  \right) + \delta \ , \text{where} \
 \delta = \begin{cases}
 \varepsilon & \text{if } \vekt{\mu}^s = {\vekt{\mu}^s}' \ , \\
 0 & \, \text{otherwise.}
 \end{cases}
\end{equation}

The hyperparameters $\mathbb{H} = \{\sigma_f, \vekt{d}, \varepsilon\}$ are determined from the training input-output pairs $\{\matr{P}_t, \vekt{y}^s_t\}$ via \emph{maximum likelihood estimation}:
\begin{equation}
\begin{split}
 \mathbb{H} 
 &= \operatorname*{argmax}_{ \hat{\mathbb{H}} } \log p(\vekt{y}^s_t|\matr{K}_{tt}(\tilde{\mathbb{H}})) 
 =
 \\
 &= \operatorname*{argmax}_{ \hat{\mathbb{H}} }
 \left\{
 -\frac{1}{2} {\vekt{y}^s_t}^\top \matr{K}_{tt}^{-1}(\hat{\mathbb{H}}) \vekt{y}^s_t
 -\frac{1}{2} \log |\matr{K}_{tt}^{-1}(\hat{\mathbb{H}})|
 -\frac{N}{2} \log (2\pi)
 \right\} \ ,
\end{split}
\end{equation}

with $ \matr{K}_{tt}(\hat{\mathbb{H}}) = k(\matr{P}_{t}, \matr{P}_{t}; \hat{\mathbb{H}}) $ being the prior covariance matrix on training data and $|\cdot|$ denoting the matrix determinant.
To this end, the box-constrained \gls{L-BFGS} optimizer with $10$ random initialization restarts is used. 
Due to standardizing outputs to unit variance (see \refsec{sec:scaling}), we can specify relatively tight bounds on the prior covariance $\sigma_f \in [10^{-2}, 10^2]$ as we expect values within few orders of magnitude around $1$. Similarly, input standardization (see \refsec{sec:feature_standardization}) allows us to set meaningful bounds for correlation lengthscales as $d_i \in [10^{-2}, 10^2]$ for any problem. The lower bound $10^{-2}$ is chosen much smaller than the average distance between the datapoints in input space and the upper bound $10^2$ is much larger than the unit standard deviation of samples in input space.
The bounds for noise are set to $\varepsilon \in [10^{-10}, 1] $.

Finally, to make predictions $\tilde{\vekt{y}}_{l,p}$ at inputs $ \matr{P}_p $ we recall the joint Gaussian distribution assumption, which also applies to realizations of training data and predictions, formally:
\begin{equation}
 \begin{pmatrix}
  \vekt{y}^s_{l,t} \\
  \tilde{\vekt{y}}^s_{l,p}
 \end{pmatrix}
 \sim
 \mathcal{N}
 \left(\vekt{0}, 
  \begin{bmatrix}
   \matr{K}_{tt} & \matr{K}_{tp} \\
   \matr{K}_{tp}^\top & \matr{K}_{pp} 
  \end{bmatrix}
 \right) \ ,
\end{equation}

where the covariance matrix blocks $ \matr{K}_{ab} = k(\matr{P}_{a}, \matr{P}_{b}) $ are determined by the prior. By conditioning the prior mean and covariance on the training data, we obtain the \emph{posterior} belief. This can be written down using the theorem of conditional Gaussian:
\begin{equation}
 p(\vekt{y}^s_{l,p} | \vekt{y}^s_{l,t}, \matr{P}_t, \matr{P}_p ) = \mathcal{N}\left( \vekt{m}_{p|t}, \matr{K}_{p|t} \right) \ ,
\end{equation}

with posterior mean $ \vekt{m}_{p|t} = \matr{K}_{pt}\matr{K}_{tt}^{-1}\vekt{y}_t $ and posterior covariance $ \matr{K}_{p|t} = \matr{K}_{pp} - \matr{K}_{pt}\matr{K}_{tt}^{-1}\matr{K}_{tp} $. Notice, that the predicted output of the regression model is the posterior mean  $ \tilde{\vekt{y}}^s_{l,p} = \tilde{\pi}^s(\matr{P}_p) = \vekt{m}_{p|t}$.
Similar to \gls{RBF} regression, again an independent function is constructed for each output dimension $l \in [1,L]$ and these gathered in the vector function $\tilde{\vekt{\pi}}^s = [\tilde{\pi}^s_1|\cdots|\tilde{\pi}^s_L]^\top$.
All functions use the anisotropic squared exponential kernel, but each is optimized separately for its hyperparameters.

For more details we refer to \cite{Dupuis2018, Kuss2006, Rasmussen2005}.
The implementation is based on the scikit-learn Python module \cite{scikit-learn}.


\subsection{Artificial neural networks}
\label{sec:ANN}

\glspl{ANN} are a computational paradigm within the field of \gls{ML} that is loosely inspired by biological neural networks. With the theoretical property of being \emph{universal function approximators} they have been used in a range of applications requiring non-linear function approximations. Different types or \emph{architectures} of \glspl{ANN} have proven to be particularly suitable in different domains, such as convolutional neural networks in image processing or recurrent neural networks in natural language processing \cite{Goodfellow-et-al-2016}. In function regression tasks,
the comparatively simple \emph{\gls{FNN}} architecture has found a lot of success \cite{Abiodun2018} and is used in this work.

As the name suggests, \gls{ANN} is a network of simple computational units called \emph{neurons}. Depending on the architecture these are arranged in different ways.
In an \gls{FNN} the neurons are organized in several \emph{layers} and each neuron is connected to all neurons in both adjacent layers. These connections are directed and information flows from the lowest to the highest layer, called \emph{input} layer $\vekt{\nu}^{i}$ and \emph{output} layer $\vekt{\nu}^{o}$, respectively. In between are multiple \emph{hidden} layers $\{ \vekt{\nu}^{h_j} \}_{1 \le j \le J} $. 
This establishes an input-output mapping
$ \tilde{\vekt{\pi}}^s: \vekt{\mu}^s \mapsto \tilde{\vekt{y}}^s, \ \tilde{\vekt{\pi}}^s(\vekt{\mu}^s; \Theta) = \tilde{\vekt{y}}^s$,
which is parametrized by the strength of the connections between neurons, described formally by the \emph{weights} ${\Theta}_W$, and \emph{biases} ${\Theta}_b$, $\Theta = \left\{ \Theta_W, \Theta_b \right\}$. Additionally, non-linear \emph{activation functions} $g$ enable non-linear behavior of the \gls{ANN}. 
Altogether an \gls{FNN} can be modeled as

\begin{equation}
\tilde{\vekt{\pi}}^s (\vekt{\mu}^s; \Theta) \
\left\{
\begin{aligned}
\vekt{\nu}^{i} &=\vekt{\mu}^s \\
\vekt{\nu}^{h_{1}} &=g^{h_{1}}\left({\Theta}_W^{h_{1}}\vekt{\nu}^{i}+{\Theta}_b^{h_{1}}\right) \\
\vekt{\nu}^{h_{j}} &=g^{h_{j}}\left({\Theta}_W^{h_{j}}\vekt{\nu}^{h_{j-1}}+\Theta_b^{h_{1}}\right) \text { for } j=2, \cdots, {J} \\
 \vekt{\nu}^{o} &=g^{o}\left({\Theta}_W^{o} \vekt{\nu}^{h_J}+{\Theta}_b^{o}\right) \\
 \tilde{\vekt{y}}^s &=\vekt{\nu}^{o} \ .
\end{aligned}
\right.
\end{equation}

In a \emph{supervised learning} paradigm, the \gls{ANN} learns the weights and biases from the presented training data.  
Although learning differs from pure optimization because ultimately the performance on unseen test data or the \emph{generalization} ability matters \cite{Goodfellow-et-al-2016}, 
the training process is posed as a non-convex optimization problem minimizing some \emph{loss function} on the training data. 
We use the standardized regression error $\varepsilon_\text{ANN}$ with \gls{RMSE} aggregation as introduced in \refsec{sec:errors}: 

\begin{equation}
 \Theta = \operatorname*{argmin}_{ \hat{\Theta} }
 \sqrt{ 1/N_{tr}
 \sum_{\vekt{\mu} \in \mathbb{P}_{tr}}
 \varepsilon_\text{ANN}^2
 \left(
 \vekt{\mu}; \tilde{\vekt{\pi}}^s_l(\vekt{\mu}^s; \hat{\Theta})
 \right) }
 =
 \operatorname*{argmin}_{ \hat{\Theta} }
 \varepsilon_\text{ANN} \left( \mathbb{P}_{tr}; \hat{\Theta} \right)
 \ .
\end{equation}

The training is typically done using gradient-based iterative optimizers
in conjuncture with \emph{backpropagation} for efficient gradient computation \cite{LeCun}. In this work, the weight update in each iteration $i$ is computed using a \emph{mini-batch} $\mathbb{P}_b \subset \mathbb{P}_{tr}$ consisting of $N_b$ elements from the training set:

\begin{equation}
    \Theta^{i+1}
    =
    \Theta^i 
    -
    \alpha G\left(\frac{\partial \varepsilon_\text{ANN} (\mathbb{P}_b; \Theta^i) }{\partial \Theta^i}\right) \ .
\end{equation}

We use $N_b = 10$ and train for $5000$ \emph{epochs} or $5000 N_{tr} / N_b$ iterations, since an epoch consists of presenting each training sample once.
$\alpha$ is the \emph{learning rate} and the function $G$ depends on the specific optimizer.
The choice of optimizer is not straight forward, since its performance can strongly depend on the choice of training hyperparameters and the problem itself \cite{Ruder2016AnOO}.
In preliminary testing, the Adam optimizer \cite{Kingma2014AdamAM} is found to perform more consistently on different problems and to converge in fewer steps than competing optimizers, such as \gls{SGD} with \gls{NAG} \cite{nesterov2013} and \gls{L-BFGS}.
Even though other optimizers perform better in specific settings, we select Adam and note, that choosing the right optimizer is an active area of research in and of itself \cite{choi2019empirical}.
The Adam hyperparameter values for stabilization and momentum decay $\epsilon=10^{-8}, \beta_1=0.9, \beta_2=0.999$ are chosen as suggested by Kingma and Ba \cite{Kingma2014AdamAM}.
Similarly to Hesthaven and Ubbiali \cite{Hesthaven2018}, we restrict ourselves to shallow \glspl{ANN} with two hidden layers $N^J=2$ each consisting of $N_\nu$ neurons. We also use the {hyperbolic tangent} activation function $g(x)=\tanh(x)= (e^x-e^{-x}) / (e^x + e^{-x})$.
Although in theory the Adam optimizer computes an adaptive learning rate for each trainable parameter, we find it beneficial to use a learning rate decay $ \alpha(e) = \alpha_0 / (1+0.005e)$, where $\alpha_0$ is the initial learning rate and $e$ is the current epoch.
It is observed that the choice of $\alpha_0$ has a significant impact on the training process, so \emph{hyperparameter-tuning} is performed for the initial learning rate $\alpha_0$ and number of neurons per hidden layer $N_\nu$. This is implemented as a grid-search over the hyperparameter space $(\alpha_0 \times N_\nu) \in [ 10^{-4}, 1 ] \times [ 10, 50 ]$ using the Tune module for Python \cite{ray2018}.
The best initial learning rate is expected to be a few orders of magnitude below that of the data, which is $\mathcal{O}(1)$ due to the standardization (see \refsec{sec:scaling}).
For an appropriate \gls{ANN}'s size we use the heuristic, that the number of learnable degrees of freedom
should be similar to the number of outputs in the training dataset $L N_{tr}$.
Using $L=20$ and $N_{tr} = 100$ (see \refsec{sec:results}) we expect the optimal $N_\nu$ to not be much larger than $35$.
A much smaller \gls{ANN} could generalize well, but might not have the flexibility to approximate the function dynamics, while a much larger \gls{ANN} might overfit.
The \gls{ANN} is implemented in PyTorch \cite{Pytorch2019}.
\section{Governing equations and numerical methods}
\label{sec:NS&solver}

In this work, we consider the fluid flow of a viscous, incompressible fluid on a time-dependent, deforming computational domain $\Omega\left(t\right) \subset \; \mathbb{R}^{n_{sd}}$, where $t \in \left[0,t_{max}\right]$ is an instant of time and $n_{sd}$ is the spatial dimension.
It is enclosed by its boundary $\Gamma\left(t\right)$. The parameter dependent quantities as the velocity $\vekt{u} = \vekt{u} \left( \vekt{x}, t ; \vekt{\mu} \right)$, pressure $p = p \left( \vekt{x}, t ; \vekt{\mu} \right)$ and temperature $T = T \left( \vekt{x}, t ; \vekt{\mu} \right)$ are governed by
incompressible Navier-Stokes and heat equations with viscous dissipation:

\begin{subequations}
\label{eq:NavierStokes}
\begin{alignat}{2}
 \nabla \cdot \vekt{u} &= 0  \quad \text{on} \ \Omega\left(t\right),
 \\
 \varrho \left( \frac{\partial \vekt{u}}{\partial t} + \vekt{u} \cdot \nabla \vekt{u} \right)
 - \nabla \cdot \vekt{\sigma} &= 0
  \quad \text{on} \ \Omega\left(t\right),
 \\
 \varrho c_p \left( \frac{\partial T}{\partial t} + \vekt{u} \cdot \nabla {T} \right)
  - \kappa \Delta {T}
  - 2 \eta \nabla\vekt{u} \colon \vekt{\varepsilon} &= 0 
   \quad \text{on} \ \Omega\left(t\right).
\end{alignat}
\end{subequations}

The fluid density $\varrho$, heat conductivity $\kappa$ and specific heat capacity $c_p$ are material specific parameters. 
The set of equations is closed by defining the Cauchy stress tensor $\vekt{\sigma}$
\begin{equation}
\label{eq:stress}
 \vekt{\sigma}(\vekt{u}, p) = -p \matr{I} + 2\eta(\dot{\gamma}, T) \vekt{\varepsilon}(\vekt{u})
 \ ,
\end{equation}

where $\vekt{\varepsilon}$ is the rate of strain tensor

\begin{equation}
\label{eq:strain}
\vekt{\varepsilon}(\vekt{u}) = \frac{1}{2}\left( \nabla\vekt{u} + \nabla\vekt{u}^\top \right)
 \ .
\end{equation}

The dynamic viscosity $\eta$ is a material constant for Newtonian fluids. Within this work we also consider generalized Newtonian fluids to model shear-thinning effects in flows of plastic melts.
In this case, the dynamic viscosity also depends on the temperature $T$ and the shear rate $\dot{\gamma} = \sqrt{2 \vekt{\varepsilon} \left( \vekt{u} \right)  \colon \vekt{\varepsilon} \left( \vekt{u} \right) }$. We use the Cross-\gls{WLF} material model to describe this relation \cite{polymer}:

\begin{subequations}
\label{eq:crossWLF}
\begin{align}
\eta \left(\dot{\gamma}, T \right)
&=
\frac{\eta_0\left(T\right) }
{1+\left( {\eta_0\left(T\right) \dot{\gamma}}/{\tau ^*} \right)^{\left(1-n\right)}},
\\
\eta_0\left(T\right) &= D_1 \; \exp \left( - \frac{A_1  \left( T - T_{ref} \right) }{A_2 + \left( T - T_{ref} \right) } \right).
\end{align}
\end{subequations}
$\tau ^*$ is the critical shear stress at the transition from the Newtonian plateau, $D_1$ is the viscosity at a reference temperature $T_{ref}$ and $A_1$ and $A_2$ are parameters that describe the temperature dependency.

The Dirichlet and Neumann boundary conditions for temperature and flow are defined as:
\begin{subequations}
\begin{alignat}{2}
    \vekt{u} = \vekt{g}^f
    & \quad \text{on} \;  \Gamma ^ f _g \left( t \right),
    \\
    \vekt{n} \cdot \vekt{\sigma} = \vekt{h}^f
    & \quad \text{on} \;  \Gamma ^ f _h \left( t \right),
    \\
    T = g^T
    & \quad \text{on} \;  \Gamma ^ T _g \left( t \right),
    \\
    \vekt{n} \cdot \kappa \vekt{\nabla} T = h^T
    & \quad \text{on} \;  \Gamma ^ T _h \left( t \right).
\end{alignat}
\end{subequations}

$ \Gamma^i _g \left( t \right)$ and $  \Gamma^i _h \left( t \right)$ are complementary portions of $  \Gamma^i  \left( t \right)$, with $i=f\mbox{ (Fluid)},\;T\mbox{ (Temperature)}$.

As the solution method, we use a \gls{DSD}/\gls{SST} finite element formulation \cite{Tezduyar1992}, which constructs the weak formulation of the governing equations for the \emph{space-time} domain. This solution method naturally takes deforming domains into account.
The time interval $[0,T]$ is divided into subintervals $I_n = [t_n, t_{n+1}]$, where $n$ defines the time level.
A space-time slab $Q_n$ is defined as the volume enclosed by the two surfaces $\Omega\left(t_n\right)$, $\Omega\left(t_{n+1}\right)$ and the lateral surface $P_n$, which is described by $\Gamma\left(t_n\right)$ as it traverses $I_n$. 
The individual space-time slabs are coupled weakly in time.
First-order interpolation is used for all degrees of freedom. 
Thus, a \gls{SUPG}/\gls{PSPG} stabilization technique is used to fulfill the \gls{LBB} condition \cite{donea2003finite}. 
For a more detailed description of the solution method, we refer to \cite{Tezduyar1992,pauli2017stabilized,Helmig2019}. \\
For the selection of snapshots for the \gls{ROM} we have to take into account that two spatial solution fields exist at a discrete time instance $t_{n+1}$: the upper field of the space-time slabs at $I_n$ and the lower field of the space-time slabs at $I_{n+1}$. These fields do not necessarily match exactly since they are only coupled weakly. However, we are only interested in the solution at the discrete time level. Thus, we only use the upper solution field of $I_n$. 
As described in \refsec{sec:Non-intrusive reduced-basis method}, an advantage of non-intrusive \gls{ROM} is that the data can stem from any discretization method or even experimental observations.
\section{Numerical results}
\label{sec:results}

In this section, we apply our standardized non-intrusive \gls{ROM} to three different fluid flow problems of increasing complexity.
We start with a validation problem in \refsec{sec:lid}, then add time dependency, parametric material properties, and temperature dependence in \refsec{subsec:OLDC_results}.
Finally, in \refsec{sec:extruder}, we move to our main use-case -- the twin-screw extruder.
Across all problems and quantities of interest, we compare the performance of the three regression models, as well as the results of \gls{ANN}'s hyperparameter tuning.

\subsection{Skewed lid-driven cavity}
\label{sec:lid}

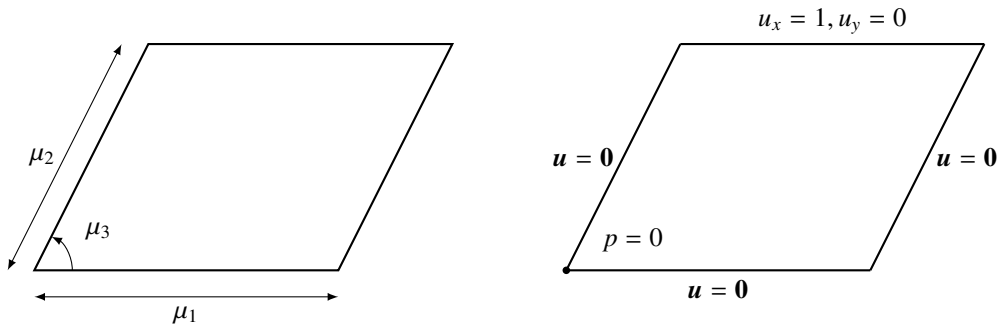
\begin{figure}[!htbp]
\centering
\begin{tikzpicture}
\tikzset{>=latex}
\def \W {4}
\def \H {3}
\def \a {60}
\def \X {7} 

\coordinate (A1) at ( 0, 0 );
\coordinate (B1) at ( {\H*cos(\a)}, {\H} );
\coordinate (C1) at ( {\H*cos(\a)+\W}, {\H} );
\coordinate (D1) at ( \W, 0 );

\draw[thick] (A1) -- (B1) -- (C1) -- (D1) -- cycle;

\pic [draw, ->, "$\mu_3$", angle eccentricity=2.0, thin] {angle = D1--A1--B1};

\draw [<->] ($(A1)+(0,-1em)$) -- ($(D1)+(0,-1em)$) node [label={[midway,below]:$\mu_1$} ] {};
\draw [<->] ($(A1)+(-1em,0)$) -- ($(B1)+(-1em,0)$) node [label={[midway,left]:$\mu_2$} ] {};

\coordinate (A2) at ($(A1)+(\X,0)$);
\coordinate (B2) at ($(B1)+(\X,0)$);
\coordinate (C2) at ($(C1)+(\X,0)$);
\coordinate (D2) at ($(D1)+(\X,0)$);

\draw[thick] (A2) -- (B2) node [label={[pos=.5, left]: $\vekt{u}=\vekt{0}$}] {};
\draw[thick] (B2) -- (C2) node [label={[pos=.5, above]: ${u_x=1, u_y=0}$} ]{};
\draw[thick] (C2) -- (D2) node [label={[pos=.5, right]: $\vekt{u}=\vekt{0}$}] {};
\draw[thick] (D2) -- (A2) node [label={[pos=.5, below]: ${\vekt{u}=\vekt{0}}$}]{};

\node at (A2)[circle,fill, inner sep=1.0pt, label={[above, right, shift={(1em,1em)}]:$p=0$}]{};

\end{tikzpicture}
\caption[Geometry and boundary conditions of the skewed lid-driven cavity.]{Parametrized geometry of the domain (left) and boundary conditions (right) of the skewed lid-driven cavity problem.}
\label{fig:lid:setup}
\end{figure}

As a validation case, we consider the skewed lid-driven cavity problem as described by Hesthaven and Ubbiali \cite{Hesthaven2018}.
%
The problem setup is shown in \reffig{fig:lid:setup}.
The computational domain consists of a parallelogram-shaped cavity. In terms of boundary conditions, no-slip conditions are imposed on the bottom and the side walls, and unit velocity at the top wall.
The pressure is fixed at zero at the lower left corner.
The parameter space for the \gls{ROM} is spanned by three geometrical parameters: horizontal length $\mu_1 \in [1,2]$, wall length $\mu_2 \in [1,2]$ and slanting angle $ \mu_3 \in [\pi/6, 5\pi/6] $.
We are interested in the steady velocity and pressure distributions, so any temperature and time effects in \refeqn{eq:NavierStokes} are neglected.
A Newtonian fluid model is used with unit density. The constant dynamic viscosity is computed depending on the geometry such that Reynolds number is always $400$ according to the dimensionless equation $ Re = {\max\{\mu_1,\mu_2\}} / {\eta} $.

A regular structured computational mesh with $100 \times 100$ nodes and space-time elements with 8-nodes is used.
Training $\mathbb{P}_{tr}$, validation $\mathbb{P}_{va}$ and test $\mathbb{P}_{te}$ sets of sizes $N_{tr}=100$, $N_{va}=50$ and $N_{te}=50$, are sampled from the parameter-space using randomized \gls{LHS}. 
A typical solution is shown in \reffig{fig:lid_examples}.
Two separate reduced bases are constructed -- one for the velocity $\vekt{u}$ and one for the pressure $p$. Both utilize $ L_{\vekt{u}} = L_{p}=20 $ basis vectors.
In each case, the three regression models described in \refsec{sec:Regression models} are trained to approximate the mapping from the standardized parameters to the standardized reduced coefficients $\tilde{\vekt{\pi}}^s : \vekt{\mu}^s \mapsto \vekt{y}^s$. The training set is used to determine the parameters of \gls{RBF} and \gls{ANN} and the hyperparameters of the \gls{GPR}. The validation set is used only for tuning the \gls{ANN}'s hyperparameters. Finally, the test set is used to quantify how the models perform on unseen data.

\def\H{30mm}
\begin{figure}[!htbp]
\begin{center}
\begin{tabular*}{\textwidth}{c @{\extracolsep{\fill}} cc}
  \includegraphics[height=\H]{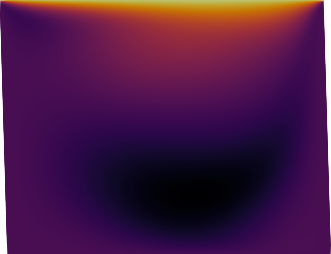} &   \includegraphics[height=\H]{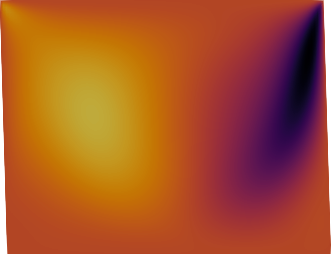} &   \includegraphics[height=\H]{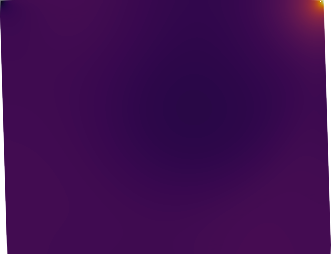} \\
  \scaleNamedInferno{$-0.37$}{$1.00$}{$u_x$} &
  \scaleNamedInferno{$-0.61$}{$0.36$}{$u_y$} &
  \scaleNamedInferno{$-0.34$}{$1.10$}{$p$}
\end{tabular*}
\end{center}
\caption[An exemplary solution of the skewed lid-driven cavity.]{An exemplary solution of the skewed lid-driven cavity for $\vekt{\mu}=(1.90,\allowbreak 1.50,\allowbreak 1.60)$.}
\label{fig:lid_examples}
\end{figure}

\begin{figure}[!htbp]
\centering
\includegraphics{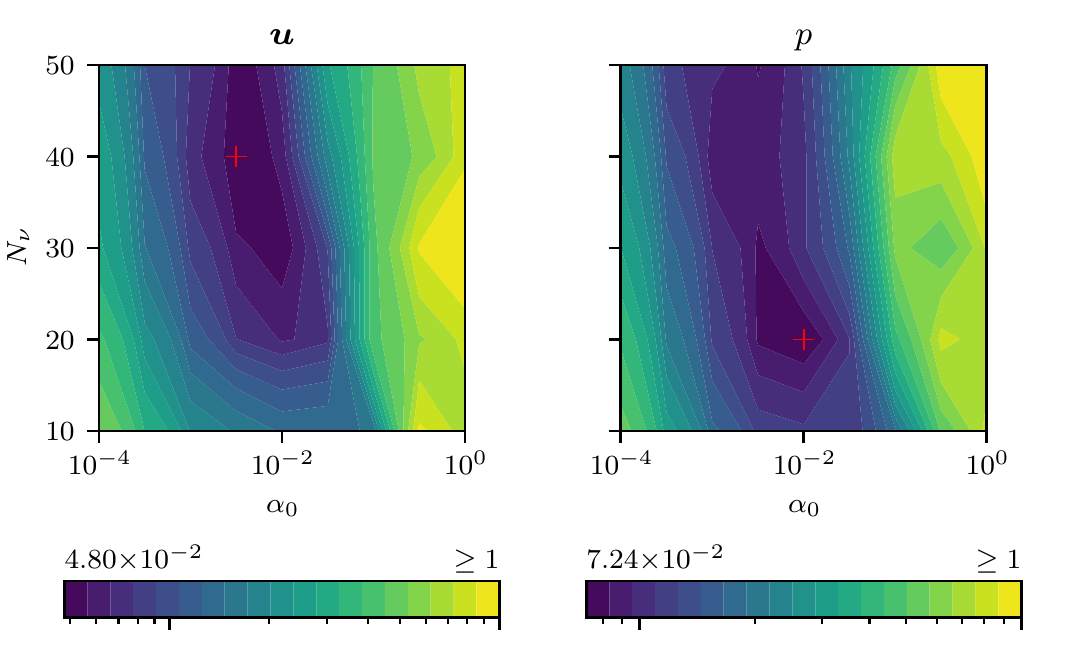}
\caption[{ANN} hyperparameter tuning results for the skewed lid-driven cavity.]{\gls{ANN} hyperparameter tuning results for the skewed lid-driven cavity. Depicted are response surfaces of aggregated standardized regression errors on the validation set $\varepsilon_\text{ANN}(\mathbb{P}_{va})$ over the initial learning rate $\alpha_0$ and number of hidden neurons $N_\nu$. The respective best configurations are marked with a cross.}
\label{fig:lid_sens}
\end{figure}

As described in \refsec{sec:ANN}, for the \gls{ANN} we first need to find an appropriate initial learning rate $\alpha_0$ and the number of neurons in both hidden layers $N_\nu$. This is implemented as a grid-search over these two hyperparameters, the results of which are illustrated in \reffig{fig:lid_sens}. 
For both parameters, the performance is less sensitive to the size of the \gls{ANN} than to the initial learning rate -- for any given $N_\nu$ an appropriate $\alpha_0$ with the performance close to the optimum can be found. This suggests that perhaps tuning only over $\alpha_0$ is a viable alternative.
Furthermore, we found that very large or very small initial learning rates lead to poor results.
Lastly, larger \glspl{ANN} tend to train better with smaller learning rates.

\begin{figure}[!htbp]
\centering
\includegraphics{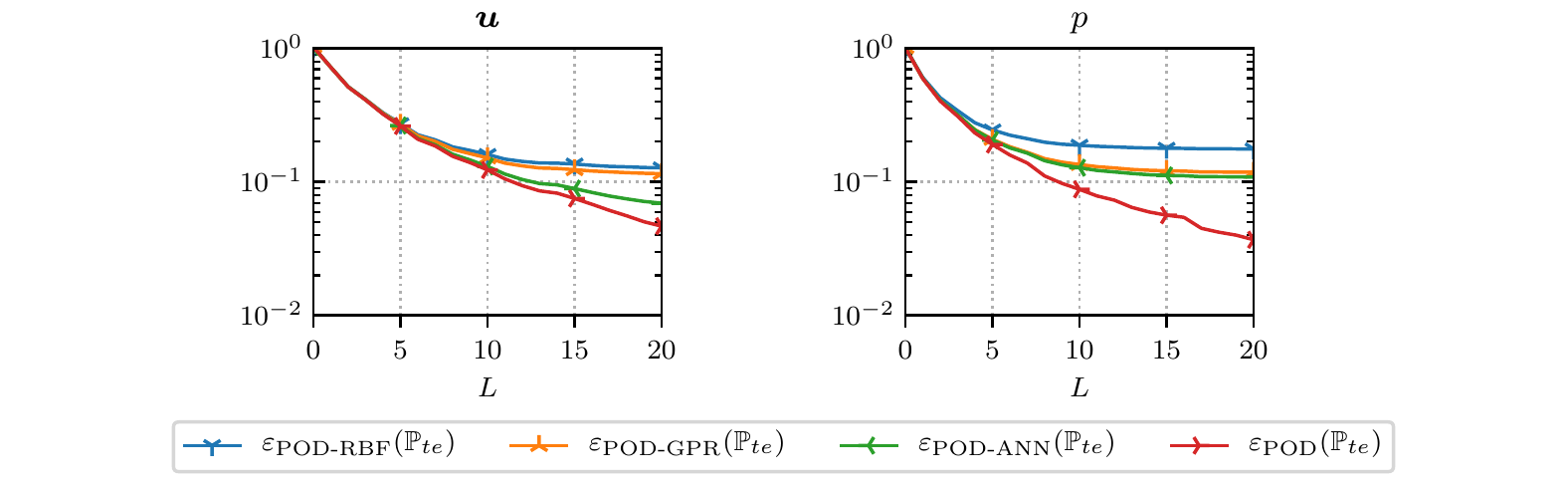}
\caption[Error analysis of different regression models for skewed lid-driven cavity.]{Error analysis of different regression models for the skewed lid-driven cavity problem. The best identified \gls{ANN} configurations marked in \reffig{fig:lid_sens} are used.}
\label{fig:lid_errors}
\end{figure}

\reffig{fig:lid_errors} compares the best identified \gls{ANN} configurations (as indicated in \reffig{fig:lid_sens}) to \gls{RBF} and \gls{GPR} in terms of the total error on the test set $\varepsilon_\text{POD-REG}(\mathbb{P}_{te})$.
The models rank as one might expect-- more flexible models perform better, i.e., \gls{ANN} outperforms \gls{GPR} which in turn outperforms \gls{RBF}.
The stagnation beyond a certain number of basis vectors $L$ is consistent with observations by other authors \cite{Graessle2019, Hesthaven2018}.

\begin{figure}[!htbp]
\centering
\includegraphics{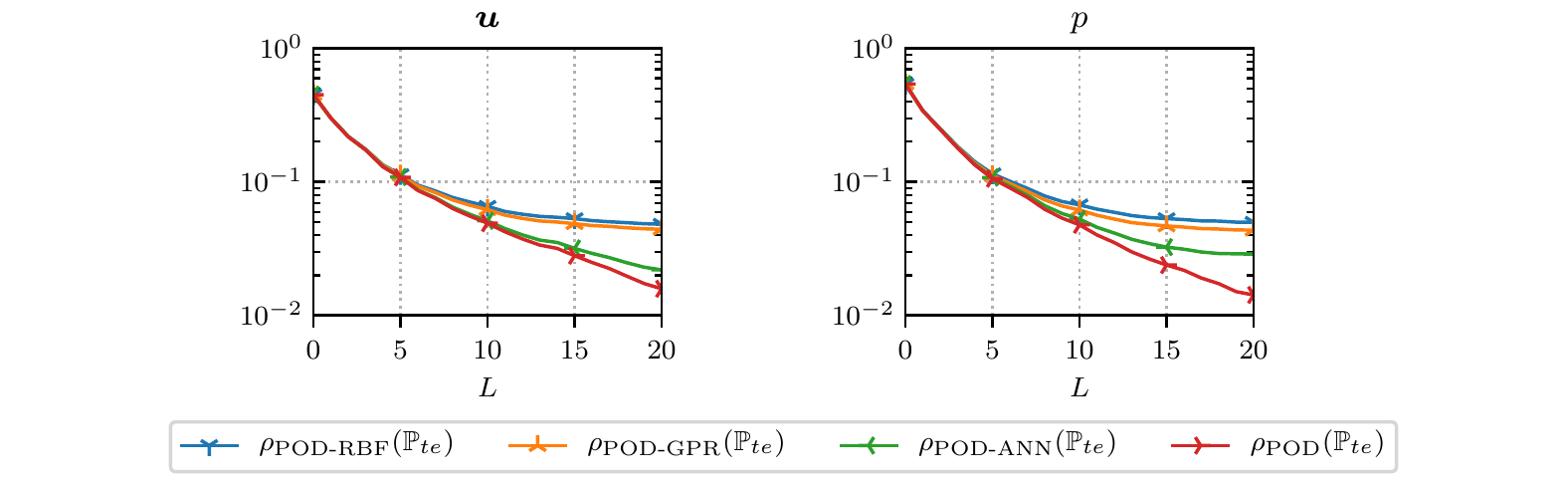}
\caption[Error analysis of different regression models for skewed lid-driven cavity.]{ \gls{ROM} performance over number of basis functions in terms of the total relative error as defined by Hesthaven and Ubbiali \cite{Hesthaven2018}.}
\label{fig:lid_errors_Hesthaven}
\end{figure}

To compare our findings with results published by Hesthaven and Ubbiali, the model performance is plotted again in \reffig{fig:lid_errors_Hesthaven} using the error definitions \cite{Hesthaven2018}:
\begin{equation}
\label{eq:rel_error}
\begin{split}
    \rho_\text{POD} (\mathbb{P}_{te}) :=&
    \frac
    { 1 }
    { N_{te} }
    \sum_{\vekt{\mu} \in \mathbb{P}_{te}}
    \frac
    { || \vekt{s} (\vekt{\mu}) - \vekt{s}_L(\vekt{\mu})) ||_2 }
    { || \vekt{s} (\vekt{\mu}) ||_2 } \ ,
    \\
    \rho_\text{POD-REG} (\mathbb{P}_{te}) :=&
    \frac
    { 1 }
    { N_{te} }
    \sum_{\vekt{\mu} \in \mathbb{P}_{te}}
    \frac
    { || \vekt{s} (\vekt{\mu}) - \tilde{\vekt{s}}_L (\vekt{\mu}) ||_2 }
    { || \vekt{s} (\vekt{\mu}) ||_2 } \ .
\end{split}
\end{equation}

In \cite{Hesthaven2018}, a similar \gls{ANN} is used with $ N_\nu^{\vekt{u}}=20 $ and $N_\nu^p=15$ neurons in both hidden layers, trained with the Levenberg-Marquardt algorithm on $100$ training snapshots.
As shown in \reftab{tab:LDC_hest}, the relative errors are the same order of magnitude as in the reference suggesting our approach and implementation is valid. 
We attribute our inferior relative POD-errors to the differences in the computed snapshots and their approximation properties, since we use a different computational mesh and numerical scheme compared to \cite{Hesthaven2018}.

\newcolumntype{L}{>{$}l<{$}} 
\def\VET{\rho_\text}
\def\E{{\times}10^}
\begin{table}[H]
\centering
\begin{tabular}{@{}lLLLL@{}}
\toprule
                                            & \multicolumn{2}{L}{\vekt{u}} & \multicolumn{2}{L}{p}      \\
\midrule
                                            & \VET{POD} & \VET{POD-ANN} & \VET{POD} & \VET{POD-ANN}     \\
This work                                   & 0.016     & 0.022         & 0.014     & 0.029             \\
Hesthaven and Ubbiali \cite{Hesthaven2018}  & 0.010     & 0.020         & 0.004     & 0.030             \\
\bottomrule
\end{tabular}
\caption[]{The achieved relative errors are in close agreement with the results reported by Hesthaven and Ubbiali \cite{Hesthaven2018}.
}
\label{tab:LDC_hest}
\end{table}

Note, that the behaviour of the relative error (\reffig{fig:lid_errors_Hesthaven}) is similar to the standardized error (\reffig{fig:lid_errors}).
This is largely due to all quantities of the benchmark problem being normalized to $\mathcal{O}(1)$ per construction.
One noticeable difference, however, is found in the performance of the \gls{ANN} for the pressure. Whereas in the the standardized error, the performance is almost identical to \gls{GPR}, in terms of relative error, \gls{ANN} outperforms \gls{GPR}.
This is likely explained by our aggregation of standardized errors relying upon \gls{RMSE}, which weights outliers more heavily than the mean aggregation used for the relative error. This in turn suggests that the \gls{GPR} deals better with more extreme solutions in pressure than \gls{ANN}.
In \reffig{fig:lid_errors_Hesthaven}, the relative errors for $L=0$ are around $50\%$ simply due to snapshot centering. In contrast, the standardized error in \reffig{fig:lid_errors} is constructed such that naively predicting the mean of snapshots results in $100\%$ error.

Lastly, \reffig{fig:lid_streamlines} shows streamlines at the same parameters as Hesthaven and Ubbiali \cite{Hesthaven2018}, who use streamlines to visually uncover minor differences in the solutions. In all projections and predictions (using \gls{POD}-\gls{ANN} with $L=20$) the main circulation zone corresponds closely to the high-fidelity snapshots. Instead, streamlines in areas of low-velocity, in particular recirculation zones are reconstructed only partially even in projections and even less in predictions. Especially in the left-most case, the predicted recirculation zone is very poor. This is also evident in the relative error, which is twice as high as the mean on the test set.
Note, that the projections can be made almost arbitrarily close to the truths by increasing the number of bases $L$, but beyond a certain point this does not benefit the prediction as illustrated in \reffig{fig:lid_errors} and \reffig{fig:lid_errors_Hesthaven}.


\def\H{30mm}
\begin{figure}[!htbp]
\begin{center}
\begin{tabular*}{\textwidth}{m{0em} c @{\extracolsep{\fill}} c @{\extracolsep{\fill}} c} 
  \rotatebox{90}{truth}&
  \includegraphics[height=\H, align=c]{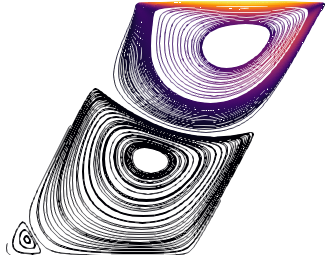} &   \includegraphics[height=\H, align=c]{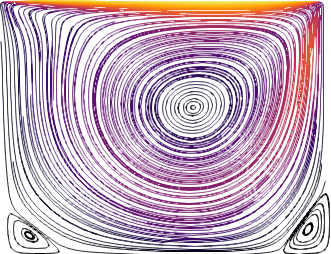} &   \includegraphics[height=\H, align=c]{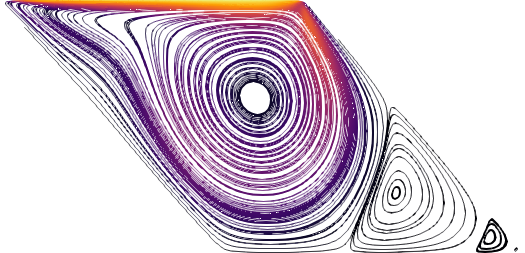} \\
   & $\vekt{\mu}=(1.12, 1.70, 1.08)$ & $\vekt{\mu}=(1.90, 1.50, 1.60)$ & $\vekt{\mu}=(1.78, 1.99, 2.29)$

 \\ \\
 \rotatebox{90}{projection} &
  \includegraphics[height=\H, align=c]{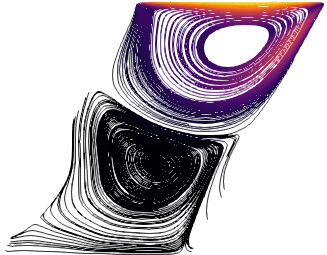} &   \includegraphics[height=\H, align=c]{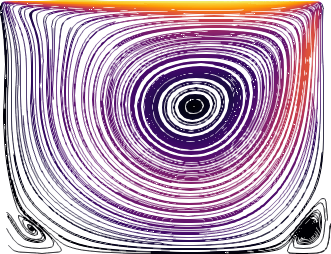} &   \includegraphics[height=\H, align=c]{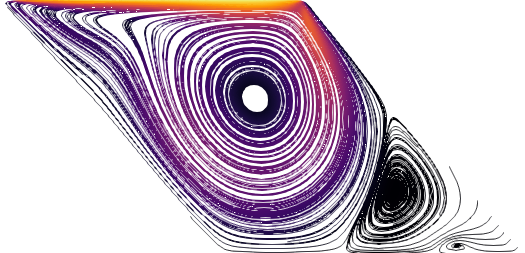} \\
  & $\varepsilon_\text{POD} = 2.3{\times}10^{-2}$ & $\varepsilon_\text{POD} = 3.6{\times}10^{-2}$ & $\varepsilon_\text{POD} = 1.7{\times}10^{-2}$ \\
  & $\rho_\text{POD} = 1.3{\times}10^{-2}$ & $\rho_\text{POD} = 1.4{\times}10^{-3}$ & $\rho_\text{POD} = 7.6{\times}10^{-3}$
  \\ \\
  \rotatebox{90}{prediction} &
  \includegraphics[height=\H, align=c]{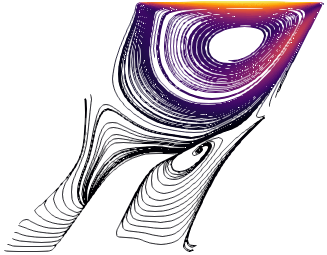} &   \includegraphics[height=\H, align=c]{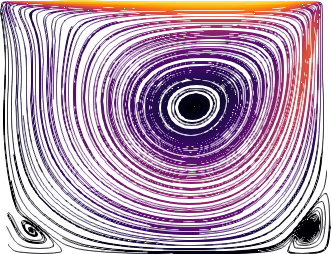} &   \includegraphics[height=\H, align=c]{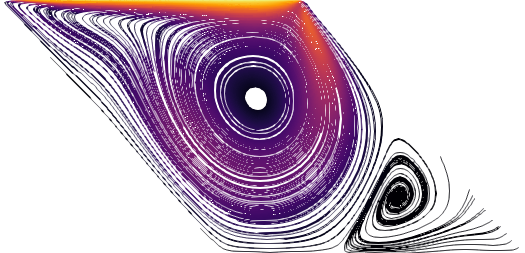} \\
  & $\varepsilon_\text{POD-ANN} = 6.8{\times}10^{-2}$ & $\varepsilon_\text{POD-ANN} = 3.8{\times}10^{-2}$ & $\varepsilon_\text{POD-ANN} = 5.2{\times}10^{-2}$ \\
  & $\rho_\text{POD-ANN} = 3.8{\times}10^{-2}$ & $\rho_\text{POD-ANN} = 1.5{\times}10^{-2}$ & $\rho_\text{POD-ANN} = 2.3{\times}10^{-2}$ \\ \\
  & & \scaleNamedInferno{$0.0$}{$1.0$}{$||\vekt{u}||$} &
\end{tabular*}
\end{center}
\caption[Streamlines for the skewed lid-driven cavity.]{Streamlines for the skewed lid-driven cavity at three different parameter values. $L=20$ bases and the \gls{ANN} regression model are used. The corresponding errors illustrate that streamlines can uncover minor differences in flows, especially in the stagnating regions.}
\label{fig:lid_streamlines}
\end{figure}

\FloatBarrier

\subsection{Oscillating lid-driven cavity}
\label{subsec:OLDC_results}

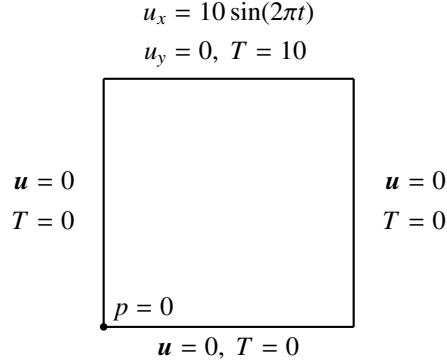
\begin{figure}[!htbp]
\centering
\begin{tikzpicture}
\def\W{0.2\textwidth}
\def\H{0.2\textwidth}
\def \a {90}

\coordinate (A1) at ( 0, 0 );
\coordinate (B1) at ( {\H*cos(\a)}, {\H} );
\coordinate (C1) at ( {\H*cos(\a)+\W}, {\H} );
\coordinate (D1) at ( \W, 0 );

\draw[thick] (A1) -- (B1) node [label={[pos=.5, right, anchor=east, shift={(-0.25,0)}]:
$\begin{aligned}
\vekt{u}&=0 \\
T &= 0
\end{aligned}$
}]{};

\draw[thick] (B1) -- (C1) node [label={[pos=.5, above]: 
$\begin{aligned}
u_x &= 10\sin(2\pi t) \\
u_y &= 0 , \
T = 10
\end{aligned}$
} ]{};

\draw[thick] (C1) -- (D1) node [label={[pos=.5, right, anchor=west, shift={( 0.25,0)}]:
$\begin{aligned}
\vekt{u}&=0 \\
T &= 0
\end{aligned}$
}]{};

\draw[thick] (D1) -- (A1) node [label={[pos=.5, below]:
$\vekt{u}=0 ,\ T = 0$
}]{};

\node at (A1)[circle,fill, inner sep=1.0pt, label={[above, right, shift={(0,.5em)}]:$p=0$}]{};

\end{tikzpicture}
\caption[Geometry and boundary conditions of the oscillating lid-driven cavity.]{Geometry and boundary conditions of the oscillating lid-driven cavity.}
\label{fig:OLDC_setup}
\end{figure}

After the comparison of our method and implementation with the literature, we now construct an intermediate problem with several characteristics found in the twin-screw extruder problem (see \refsec{sec:extruder}), namely, time dependency, parametrization of material properties and temperature distribution as output.
Similar to the first problem, the \emph{unsteady} lid-driven cavity problem is a common benchmark for time dependent flows. However, in \gls{ROM} literature often either only the velocity is examined as a quantity of interest or the problem is parametrized only in time \cite{Stabile2018, Graessle2019, Deshmukh2016, Chen2018, Balajewicz2012ANA, Dumon2013}. 
Thus we define the \emph{oscillating lid-driven cavity} problem as illustrated in \reffig{fig:OLDC_setup}.

At the top of a unit square domain, an oscillating velocity of $10\sin\left(2\pi t\right)$ is imposed in the horizontal direction along with a constant temperature of $10$. On the bottom and side walls we again impose a no-slip boundary condition for the velocity and homogeneous Dirichlet boundary conditions for the temperature. As initial condition, we chose zero everywhere for both the velocity and the temperature. We compute 100 time steps with a size of 0.01. This corresponds to a single full oscillation of the sinusoidal lid velocity.
Due to the vortices that develop in the cavity, the warm top fluid is transported into the interior forming distinct temperature traces whose characteristic depends on the heat conductivity and viscosity of the Newtonian fluid, parametrized as $\kappa = \mu_1 \in [0.001, 0.01]$ and $\eta = \mu_2 \in [0.01, 0.1]$. The fluid density and heat-capacity are fixed at $\varrho = 1$ and $c_p=1$, respectively, and again a spatial mesh of $100\times100$ nodes is used.
An exemplary solution is shown in \reffig{fig:OLDC_sols} and in the top row of \reffig{fig:OLDC_preds}. 
Notice that the temperature fields are particularly rich in features, since the advective terms are more dominant than the diffusive terms.

\def\W{0.2\textwidth}
\def\Wcs{0.15\textwidth}
\begin{figure}[!htbp]
\centering
\begin{tabular*}{\textwidth}{c @{\extracolsep{\fill}} c @{\extracolsep{\fill}} c @{\extracolsep{\fill}} c}
  \includegraphics[width=\W]{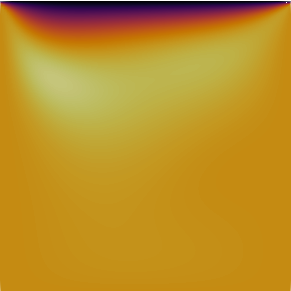} &
  \includegraphics[width=\W]{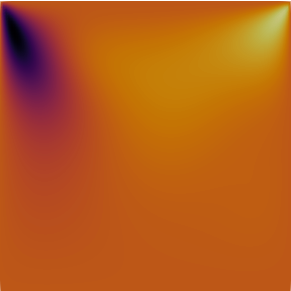} &
  \includegraphics[width=\W]{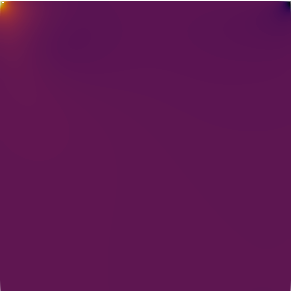} &
  \includegraphics[width=\W]{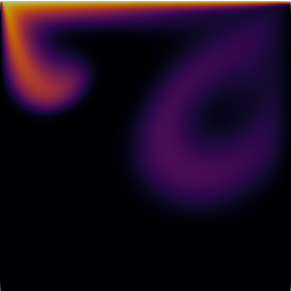} \\
  \scaleNamedInfernoWidth{$-10$}{$2.1$}{$u_x$}{\Wcs} &
  \scaleNamedInfernoWidth{$-4.9$}{$2.3$}{$u_y$}{\Wcs} &
  \scaleNamedInfernoWidth{$-69$}{$140$}{$p$}{\Wcs} &
  \scaleNamedInfernoWidth{$0$}{$10$}{$T$}{\Wcs}
\end{tabular*}
\caption[An exemplary solution of the oscillating lid-driven cavity.]{An exemplary solution of the oscillating lid-driven cavity for $\vekt{\mu}=(0.055,\allowbreak 0.0069)$ at time step $t_{75}$.}
\label{fig:OLDC_sols}
\end{figure}

\begin{figure}[!htbp]
\centering
\includegraphics{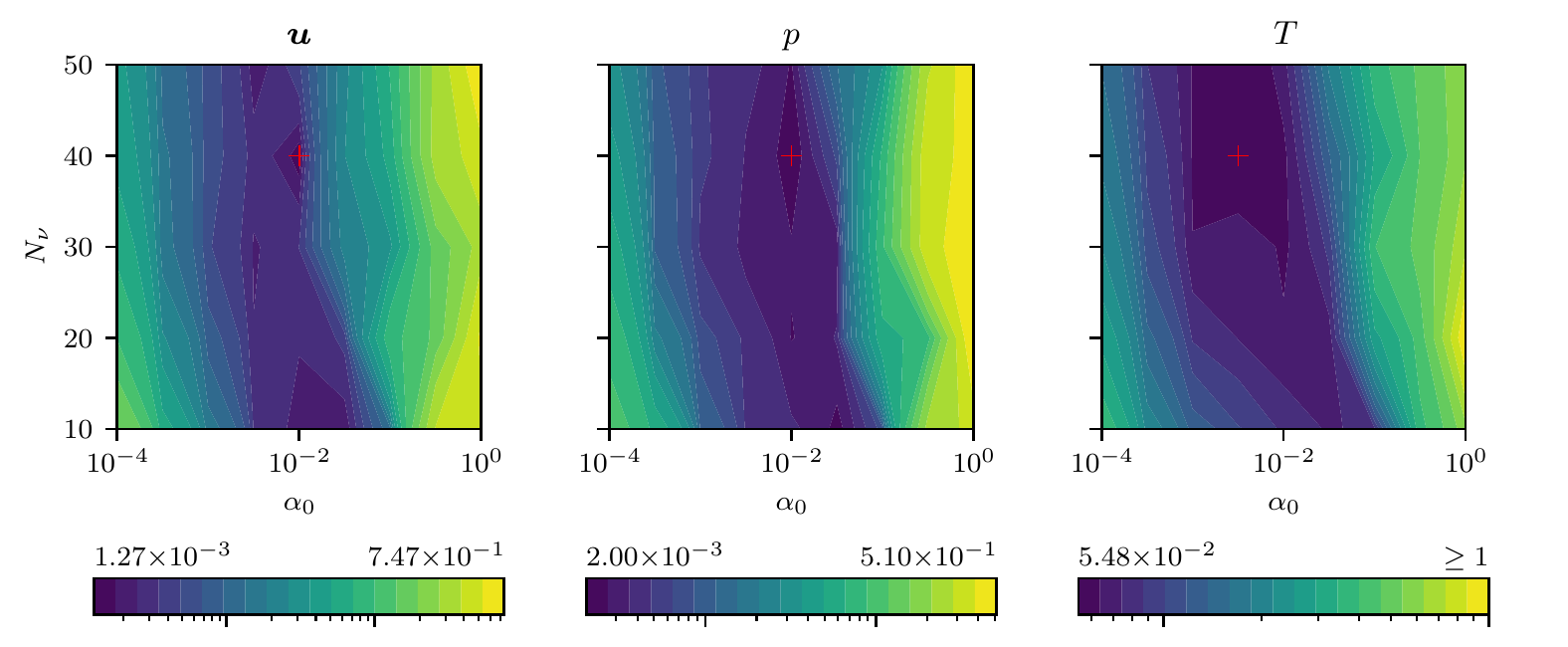}
\caption[{ANN} hyperparameter tuning results for the oscillating lid-driven cavity.]{\gls{ANN} hyperparameter tuning results for the oscillating lid-driven cavity. Depicted are response surfaces of aggregated standardized regression error on the validation set $\varepsilon_\text{ANN}(\mathbb{P}_{va})$ over the initial learning rate $\alpha_0$ and number of hidden neurons $N_\nu$. The respective best configurations are marked with a cross.}
\label{fig:OLDC_sens}
\end{figure}

To deal with the introduced time dependency when performing \gls{POD} and regression, time is treated similar to space as proposed in \refsec{sec:unsteadyRB}.
In addition to the preprocessing steps, this allows to follow the same exact procedure as described in \refsec{sec:lid} to construct and assess the \glspl{ROM}.
The results of the \gls{ANN} tuning are displayed in \reffig{fig:OLDC_sens}. The qualitative behaviour of the tuning on the same hyperparameter space closely resembles the skewed lid-driven cavity problem in \reffig{fig:lid_sens}. This illustrates the intended purpose of standardization and is discussed in more detail in \refsec{sec:extruder}. Again, the performance is much less sensitive to the size of the network than the learning rate and the optimal learning rates are similar.

\begin{figure}[!htbp]
\centering
\includegraphics{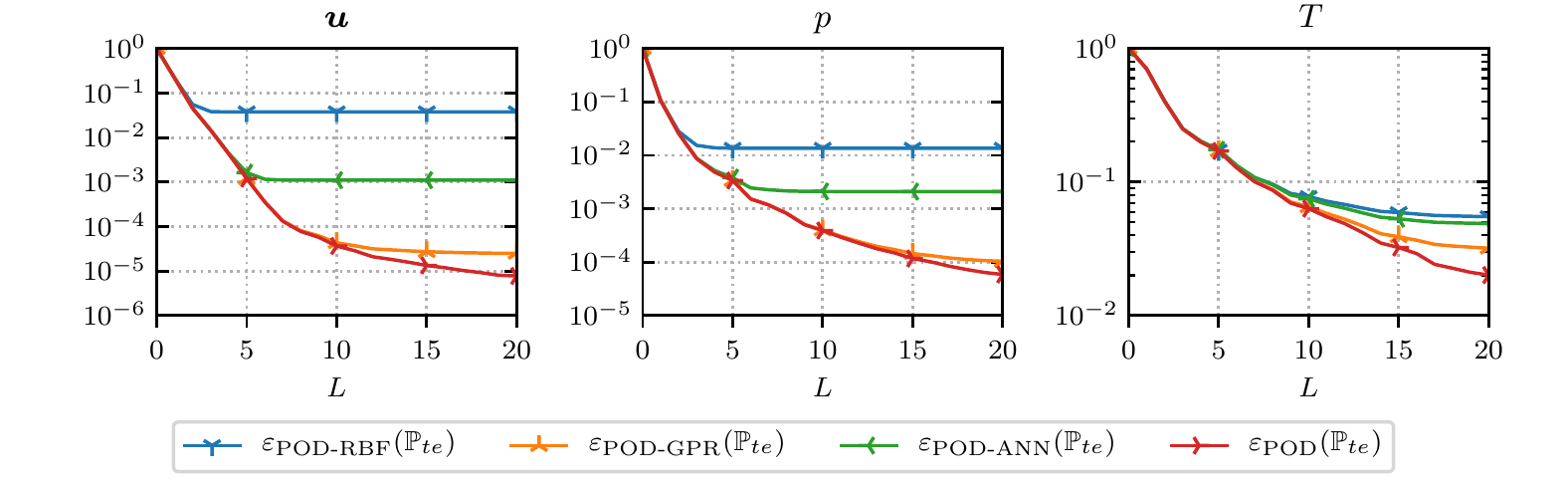}
\caption[Error analysis of different regression models for oscillating lid-driven cavity.]{Error analysis of different regression models for the oscillating lid-driven cavity problem. The best identified \gls{ANN} configurations marked in \reffig{fig:OLDC_sens} are used.}
\label{fig:OLDC_errors}
\end{figure}

\reffig{fig:OLDC_errors} depicts the errors resulting from the best \gls{ANN} configurations together with the other two regression models.
Interestingly, both projection and prediction errors for velocity and pressure are several orders of magnitude smaller than for the skewed lid-driven cavity problem. We hypothesize this is due to $\vekt{u}$ and $p$ effectively being parametrized only in viscosity $\nu = \mu_1$ because the temperature equation is decoupled from the flow equations for a Newtonian fluid (see Equations \labelcref{eq:NavierStokes,eq:stress,eq:strain,eq:crossWLF}).
Since the parameter space is sampled with \gls{LHS}, no two training parameter samples have equal viscosities $\mu_1^{(i)} \neq \mu_1^{(j)} \ \forall i \neq j$ and the effectively one-dimensional parameter space is sampled more densely.
Temperature on the other hand depends on both parameters and shows greater complexity and errors.

In this test case, the \gls{GPR} significantly outperforms the \gls{ANN}, especially for $\vekt{u}$ and $p$ despite the high computational investment associated with hyperparameter tuning.
We argue this is due to our choice of \gls{GPR}'s anisotropic kernel (see \refeqn{eq:GPR_aniso}), which can learn a unique correlation length for each input dimension.
For all bases of $\vekt{u}$ and $p$ the learned lengthscales are less than one for viscosity $d_1 < 1$ indicating the presence of local dynamics.
However, for heat conductivity most learnt lengthscales are at the upper bound of the box-constrained search space $d_2 = 100$. This implies that the anisotropic \gls{GPR} successfully recognizes the stationarity with respect to the second parameter.
Increasing the upper bound might even further improve the \gls{GPR}'s performance. We can test the limit case by manually fixing $d_2$ at infinity, i.e., explicitly neglecting the heat-conductivity. The resulting decrease in the regression errors can be seen in \reftab{tab:OLDC_iso}. Note, that the achieved regression errors are smaller than the projection errors, suggesting a very good fit and indicating that $L$ can be increased to further reduce the total error.
If instead the \emph{isotropic} squared exponential kernel (see \refeqn{eq:GPR_iso}) is used, the \gls{GPR} ranks between \gls{ANN} and \gls{RBF} ($\varepsilon^\text{iso}_\text{GPR}(\mathbb{P}_{te})=1.3{\times}10^{-2}$ and $4.3{\times}10^{-3}$ for  $\vekt{u}$ and $p$, respectively).
Despite the \gls{ANN} also having the flexibility to represent the stationarity with respect to $\mu_2$, it is not as straight-forward to investigate whether this behaviour is learnt due to the abstract role of \gls{ANN}'s weights.
Instead, we can explicitly enforce this behaviour prior to hyperparameter tuning by removing the second input neuron altogether. Even though the regression errors decrease roughly five-fold, the result is still a magnitude worse than the anisotropic \gls{GPR}.
However, this observed disadvantage of \glspl{ANN} is consistent with results in the literature. Especially relative to their computational cost, \glspl{GPR} reportedly tend to outperform \glspl{ANN} on small 
and smooth, i.e., densely sampled datasets, whereas \glspl{ANN} tend to excel at generalizing non-local function dynamics \cite{Kuss2006, Rasmussen1997}.
\Gls{RBF} regression benefits the most from ignoring the second parameter, since it is isotropic per construction. In the one-dimensional and densely-sampled setting, it performs almost as well as the \gls{GPR}. However, this demonstrates how \gls{RBF} might underperform in other problem settings with hidden anisotropy. 
 In principle, more exotic anisotropic \gls{RBF} regression methods have been proposed \cite{Casciola2007}, but have received little attention, especially in the \gls{ROM} community.
 
\newcolumntype{L}{>{$}l<{$}} 
\def\VET{\varepsilon_\text}
\def\E{{\times}10^}
\begin{table}[H]
\centering
\begin{tabular}{@{}LLLLLLL@{}}
\toprule
                                    & \multicolumn{3}{L}{\vekt{u}}      & \multicolumn{3}{L}{p}             \\
\midrule
                                    & \VET{RBF} & \VET{GPR} & \VET{ANN} & \VET{RBF} & \VET{GPR} & \VET{ANN} \\
\tilde{\vekt{\pi}}^s(\mu_1, \mu_2)  & 3.8\E{-2} & 2.4\E{-5} & 1.6\E{-3} & 1.4\E{-2} & 8.6\E{-5} & 2.0\E{-3} \\
\tilde{\vekt{\pi}}^s(\mu_1)         & 8.1\E{-6} & 5.3\E{-6} & 6.7\E{-4} & 4.7\E{-5} & 3.5\E{-5} & 1.2\E{-3} \\
\bottomrule
\end{tabular}
\caption[Regression errors when ignoring the second input parameter in the oscillating lid-driven cavity.]{Regression errors when using both input parameters versus ignoring the heat-conductivity, which the velocity and pressure do not depend on. Some regression errors are less than the respective projection errors $\varepsilon_\text{POD}=7.8{\times}10^{-6}$ and $5.9{\times}10^{-5}$ for $\vekt{u}$ and $p$, suggesting a very good fit.
}
\label{tab:OLDC_iso}
\end{table}

\reffig{fig:OLDC_preds} illustrates the true temperature distribution of a sample from the test set alongside the absolute error of the predicted solution made by the \gls{GPR} over several time steps. As expected, the largest errors manifest in the most dynamic regions.

\def\W{0.2\textwidth}
\def\Hcs{0.15\textwidth}
\begin{figure}[!htbp]
\centering
\begin{tabular*}{\textwidth}{l @{\extracolsep{\fill}} c @{\extracolsep{\fill}} c @{\extracolsep{\fill}} c @{\extracolsep{\fill}} c}
  \scaleNamedInfernoHeight{$0$}{$10$}{$T$}{\Hcs} &
  \includegraphics[width=\W]{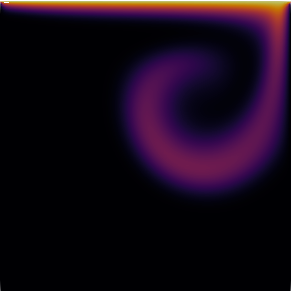} &
  \includegraphics[width=\W]{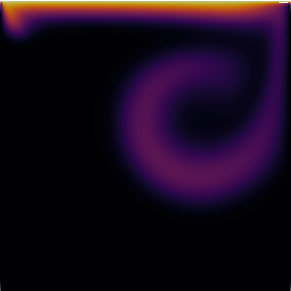} &
  \includegraphics[width=\W]{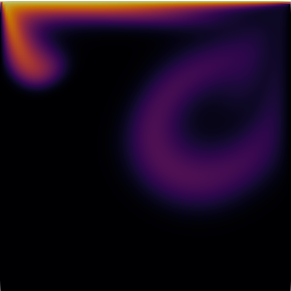} &
  \includegraphics[width=\W]{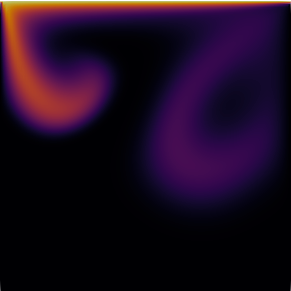} \\
  \scaleNamedInfernoHeight{$-0.032$}{$0.032$}{$T-\tilde{T}$}{\Hcs} &
  \includegraphics[width=\W]{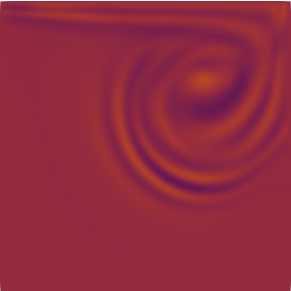} &
  \includegraphics[width=\W]{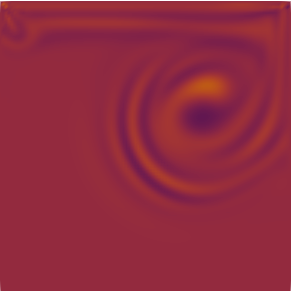} &
  \includegraphics[width=\W]{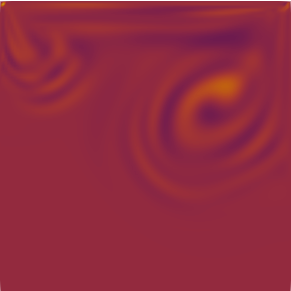} &
  \includegraphics[width=\W]{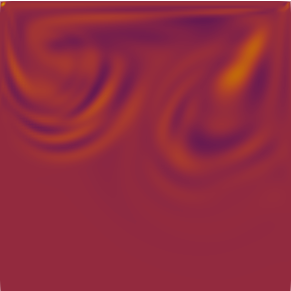} \\
  & $t_{50}$ & $t_{60}$ & $t_{70}$ & $t_{80}$ \\ 
\end{tabular*}
\caption[True temperature distribution and error in prediction over several time steps for oscillating lid-driven cavity.]
{The high-fidelity solution $T(\vekt{\mu}=(0.055,\allowbreak 0.0069) \in \mathbb{P}_{te})$ (top) and the error in prediction made by the best \gls{GPR} (bottom) at several time steps. Over all investigated time steps the standardized total error in the prediction is $\varepsilon_\text{POD-GPR}=3.9{\times}10^{-3}$.
}
\label{fig:OLDC_preds}
\end{figure}

\begin{figure}[!htbp]
\centering
\includegraphics{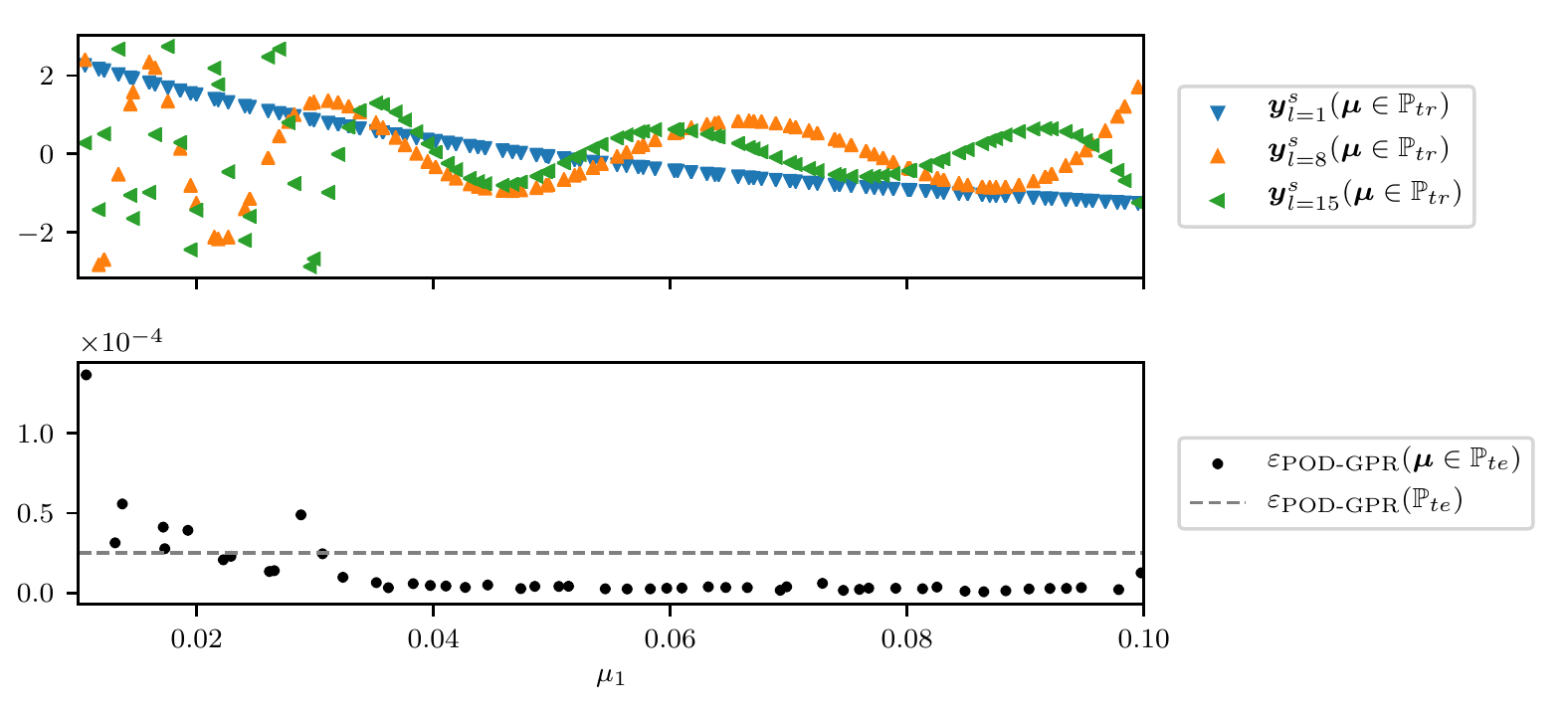}
\caption[Standardized reduced coefficients and prediction errors for velocity over viscosity for the oscialling lid-driven cavity problem.]{A few select standardized reduced coefficients of the training set $\mathbb{P}_{tr}$ over the viscosity $\nu=\mu_1$ for the velocity $\vekt{u}$ for the oscillating lid-driven cavity problem (top). For small $\mu_1$ the higher regression maps are undersampled which leads to large prediction errors in this region (bottom).}
\label{fig:OLDC_eps_over_viscosity}
\end{figure}

Notice, that the error of the particular prediction $\varepsilon_\text{POD-GPR}(\vekt{\mu}=(0.055,\allowbreak 0.0069) \in \mathbb{P}_{te}) = 3.9{\times}10^{-3} $ 
is an order of magnitude smaller than the aggregated error over the whole test set ${\varepsilon}_{POD-GPR}(\mathbb{P}_{te}) = 3.2{\times}10^{-2}$.
Due to \gls{RMSE} aggregation, a few outliers can determine the magnitude of the aggregated error. In our case, this manifests by errors not being distributed evenly over the parameter space. This is best illustrated on the effectively one-dimensionally parametrized velocity $\vekt{u}$. \reffig{fig:OLDC_eps_over_viscosity} shows a few training samples of the underlying regression maps which tend to become more dynamic towards small viscosities. For the first bases this effect is negligible, but as $l$ increases towards $L$, this region becomes undersampled and has higher test errors. Bottom row of \reffig{fig:OLDC_eps_over_viscosity} illustrates this for $\varepsilon_\text{POD-GPR}\left(\mathbb{P}_{te}\right)$, but all other investigated models, as well as the projection error behave the same. This is an indication that \emph{adaptive sampling methods} can offer a performance boost \cite{Yondo2018}.

\FloatBarrier
\subsection{Twin-Screw Extruder}
\label{sec:extruder}

For our main use-case, we consider the flow of a plastic melt inside a cross-section of a twin-screw extruder. It is characterized by a time- and temperature-dependent flow of a generalized Newtonian fluid on a moving domain. 
In contrast to previous problems, the non-Newtonian Cross-\gls{WLF} fluid model is used for which the viscosity $\eta(\dot{\gamma}, T)$ depends on both the shear rate and temperature (see \refeqn{eq:crossWLF}). This constitutes a strongly coupled problem, in which not only the flow influences the temperature distribution but also vice-versa. This problem has been investigated in a non-parametrized and non-\gls{ROM} context in \cite{hinz2020boundary}.
As mentioned in \refsec{sec:Introduction}, the moving domain and the small gap sizes make the problem numerically challenging. 
Thus, the computational burden in a many-query context is very high and the potential of non-intrusive \gls{ROM} is promising. In this preliminary work, we restrict the investigation to a two-dimensional cross-section of the extruder.

\begin{figure}[!htbp]
\centering
\input{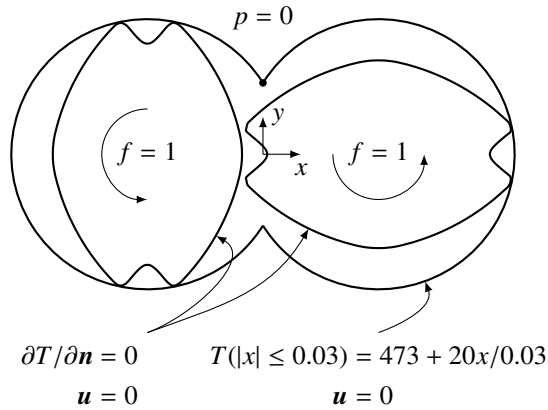}
\caption[Geometry and boundary conditions of the twin-screw extruder.]{Geometry and boundary conditions of the twin-screw extruder.}
\label{fig:EXTR_setup}
\end{figure}

The problem setup in \reffig{fig:EXTR_setup} depicts two screws rotating with unit frequency $f$ inside a barrel with the geometries and material properties used in \cite{hinz2020boundary}.
On all boundaries the no-slip condition is prescribed to the fluid.
For the temperature, we impose a Dirichlet boundary condition on the barrel that increases along the x-direction from 453 to 493; the screws are assumed to be adiabatic.
The pressure is fixed at zero at the upper cusp point. We are compute $100$ time steps of size $0.0125$, in total corresponding to $1 \frac{1}{4}$ full revolutions.
We use a structured spatial mesh with $2459$ nodes. The mesh is adapted to the changing domain over time using a spline-based mesh update technique \cite{hinz2020boundary} based on the \emph{\gls{SRMUM}} \cite{Helmig2019}.
The initial fluid temperature is $473$ and velocities are zero. All quantities can be interpreted in their respective SI units.
The constants in \refeqn{eq:crossWLF} describing the Cross-\gls{WLF} fluid are $A_1=28.315$, $A_2=51.6$, $T_{ref}=263.14$, $\tau^*=25000$ and $n=0.2923$. The specific heat capacity is $c_p=1000$.
The screw movement forces the fluid through the small gap between screws with a large velocity and creates a significant pressure drop. The gap region is subject to large shear rates and thus local differences in viscosity. Due to viscous dissipation, the fluid heats up and is transported upwards with the left screw and downwards with the right. This can be seen in exemplary solutions in \reffig{fig:extr_sols} and the left column of \reffig{fig:extr_truth_pred}.
To build the \gls{ROM}, we consider a parameter space spanned by thermal conductivity $ \kappa = \mu_1 \in [10^0, 10^1]$, reference viscosity $D_1 = \mu_2 \in [10^{12},10^{13}]$ and density $\varrho = \mu_3 \in [10^2, 10^3]$ and again generate $100$, $50$ and $50$ high-fidelity solutions for training, validation and testing, respectively. $L=20$ bases are used for velocity, pressure and temperature alike.

\def\W{0.42\textwidth}
\begin{figure}[!htbp]
\begin{center}
\begin{tabular*}{\textwidth}{c @{\extracolsep{\fill}} c}
  \includegraphics[width=\W]{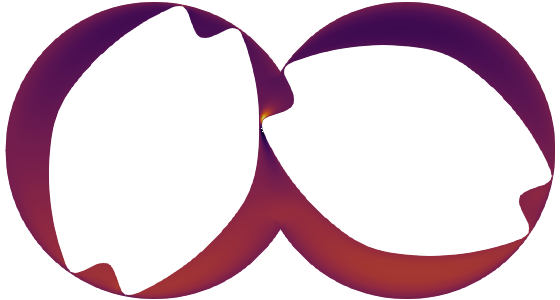} &
  \includegraphics[width=\W]{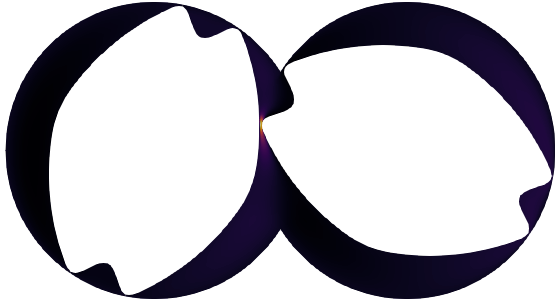} \\ \rule{0pt}{6ex}  
	\scaleNamedInferno{$-0.25$}{$0.37$}{$u_x$} &
	\scaleNamedInferno{$-0.1$}{$1.3$}{$u_y$} \\
	\\
  \includegraphics[width=\W]{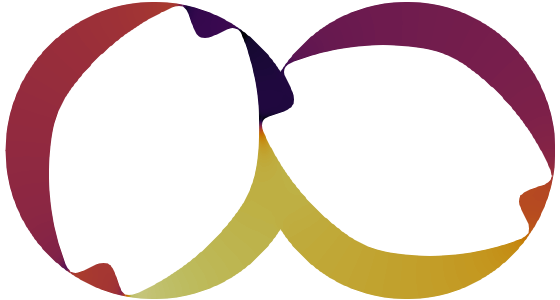} &  
  \includegraphics[width=\W]{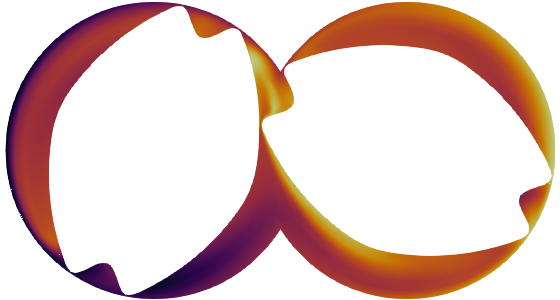} \\  \rule{0pt}{6ex} 
  \scaleNamedInferno{$-5.7{\times}10^5$}{$3.3{\times}10^6$}{$p$} &
  \scaleNamedInferno{454}{493}{$T$} \\
\end{tabular*}
\label{fig:extr_sol}
\end{center}
\caption[Exemplary solution of the twin-screw extruder.]{Exemplary solution of the twin-screw extruder for $\vekt{\mu}=(1.01,\allowbreak 6.52{\times}10^{12},\allowbreak 504)$ at time step $t_{75}$.}
\label{fig:extr_sols}
\end{figure}

\begin{figure}[!htbp]
\centering
\includegraphics{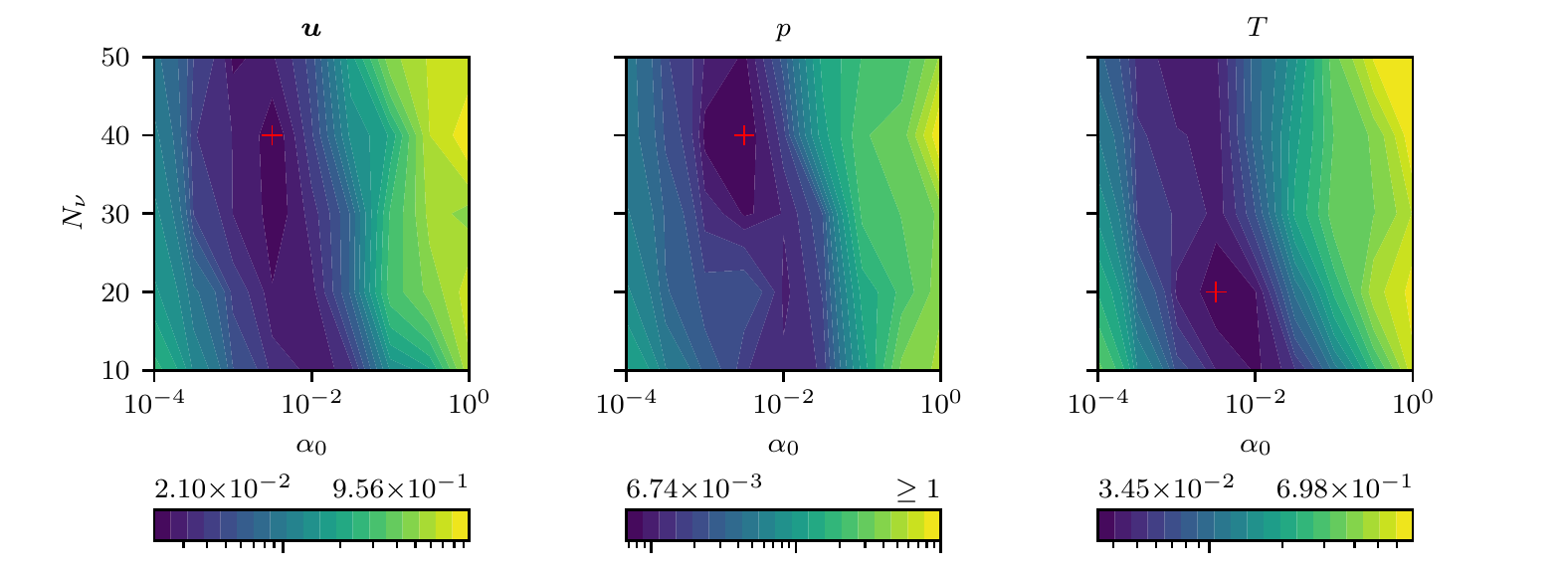}
\caption[{ANN} hyperparameter tuning results for twin-screw extruder.]{\gls{ANN} hyperparameter tuning results for twin-screw extruder. Depicted are response surfaces of aggregated standardized regression error on the validation set $\varepsilon_\text{ANN}(\mathbb{P}_{va})$ over the initial learning rate $\alpha_0$ and number of hidden neurons $N_\nu$. The respective best configurations are marked with a cross.}
\label{fig:extr_sens}
\end{figure}

This more practical problem illustrates clearly, why input scaling (as introduced in \refsec{sec:feature_standardization}) is necessary. Due to the vastly different parameter magnitudes, the \gls{RBF} would otherwise effectively neglect $\mu_1$ and $\mu_3$ when computing the distance in parameter space. Although the anisotropic \gls{GPR} has the ability to learn different lengthscales, this would require to manually specify problem-dependent search bounds for each parameter. 
\glspl{ANN} also show faster convergence with standardized inputs \cite{LeCun98}.
Similarly, the temperature distribution motivates the use of standardized errors and snapshot centering. If, for example, the relative error as in \refeqn{eq:rel_error} was used, naively predicting the mean flow or even a constant temperature of e.g. $473$ would result in deceptively small relative errors due to the variation in temperature being much smaller than the offset. This could be remedied by re-stating the problem in a normalized manner. However, the standardized errors together with preprocessing steps achieve the same effect; moreover, this is also more in line with the 'data-driven' approach. This allows to easily apply the same presented workflow to any problem or data with the results being intuitive, comparable across problems and independent of choice of units. 
This can be seen in the results of the \gls{ANN} hyperparameter tuning in \reffig{fig:extr_sens}, which are once again similar to both previous problems (see \reffig{fig:lid_sens} and \reffig{fig:OLDC_sens}) despite the very different problem setting.
As in all other cases, the optimal learning rate is between $10^{-3}$ and $10^{-2}$ and the number of hidden units is between $20$ and $40$. The results correspond well to the heuristic used in \refsec{sec:ANN}, even though the error is less sensitive to the size.

\begin{figure}[!htbp]
\centering
\includegraphics{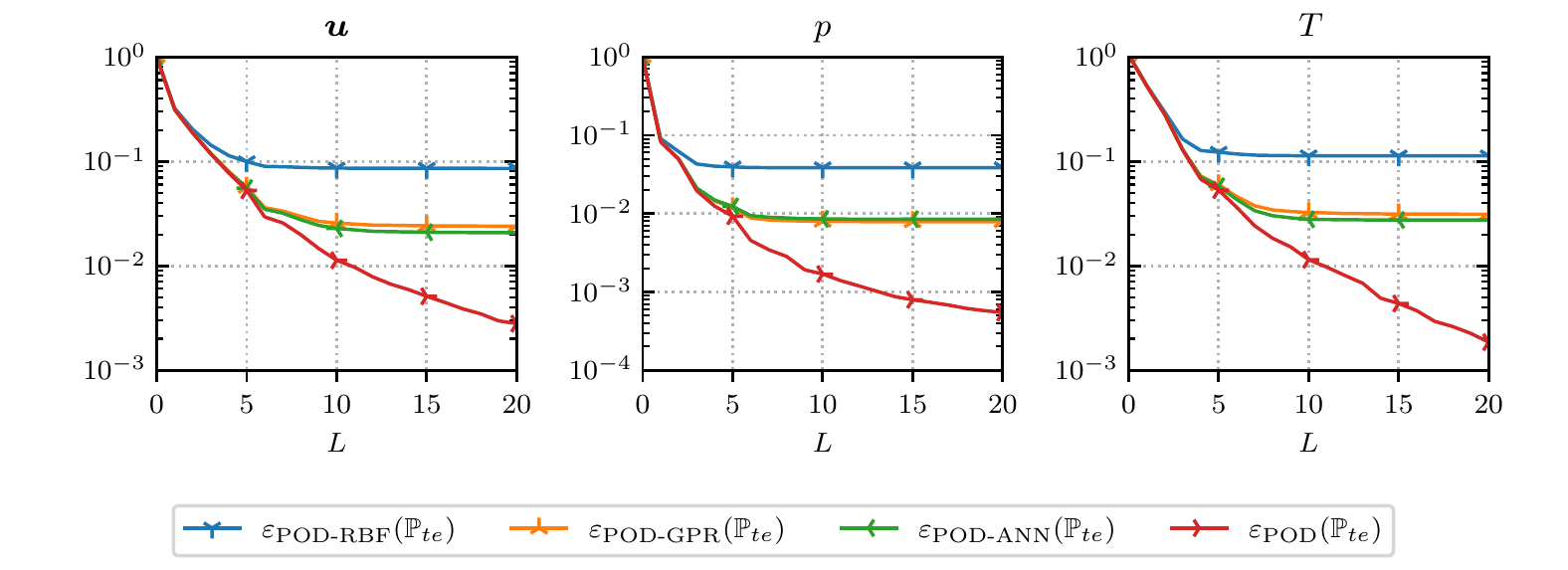}
\caption[Error analysis of different regression models for the twin-screw extruder problem.]{Error analysis of different regression models for the twin-screw extruder problem using $N_{tr}=100$ training samples. The best identified \gls{ANN} configurations marked in \reffig{fig:extr_sens} are used.}
\label{fig:extr_errors}
\end{figure}

\reffig{fig:extr_errors} depicts the best \gls{ANN} configurations together with the other two regression models.
Again, the tuned \gls{ANN} does \emph{not} outperform the \gls{GPR}
despite the computationally expensive hyperparameter tuning. Furthermore, the greater flexibility of the \gls{ANN} also requires making several decisions about the fixed and tuned hyperparameters, which usually boils down to empirical and tedious trial-and-error.

Bear in mind, however, that our results suggest that these manual efforts can be avoided and an \gls{ANN} can be trained with good success by following the described preprocessing steps, using the standardized error, and automated hyperparameter search. Nevertheless, in all the investigated settings \gls{GPR} is at least as good as \gls{ANN}, despite it being easier to implement, train and interpret.

For this problem, the performance of the \gls{RBF} is significantly worse than in the other two problems. Even so, since it can be implemented and trained in a fraction of time even compared to \gls{GPR}, \gls{RBF} regression can serve as a viable empirical upper-bound for the other models.

\begin{figure}[!htbp]
\centering
\includegraphics{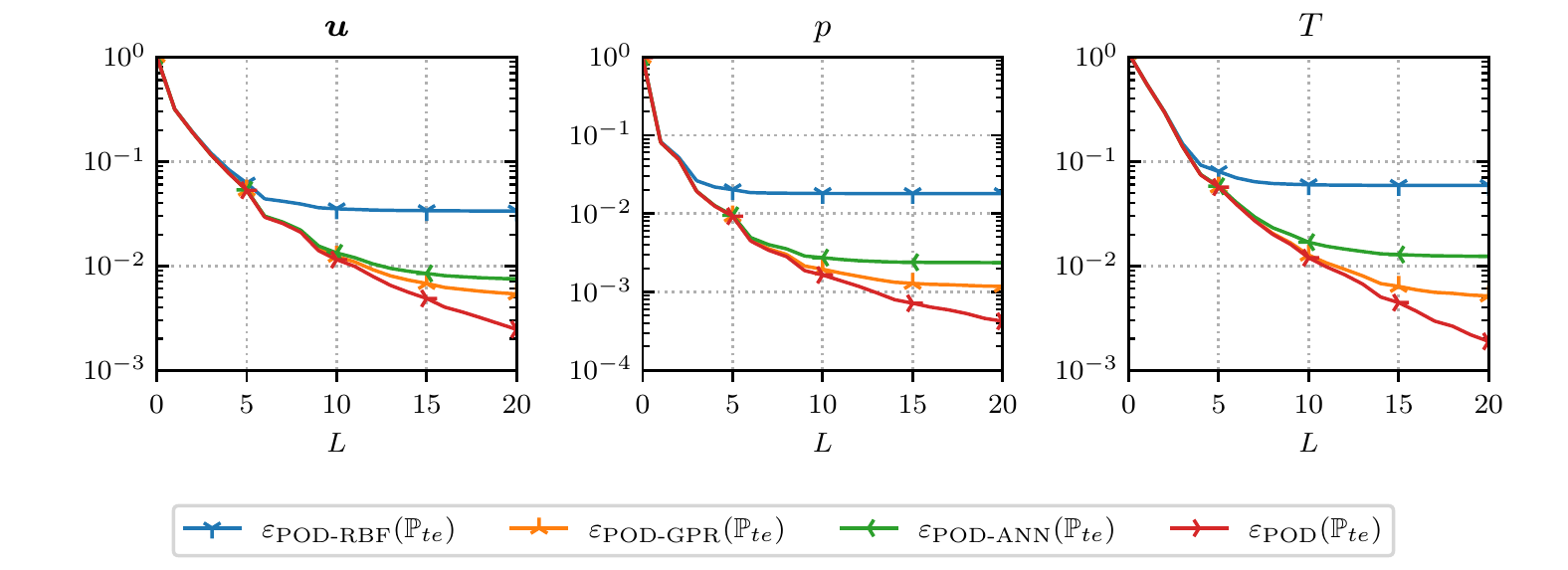}
\caption[]{Error analysis for the twin-screw extruder problem using $N_{tr}=400$ training samples. The best identified \gls{ANN} configurations described in the text are used.}
\label{fig:extr_400_errors}
\end{figure}

To investigate the nature of errors, \reffig{fig:extr_truth_pred} illustrates the true temperature over several timesteps alongside the absolute error of the predicted solution computed using the \gls{POD}-\gls{ANN}. As expected, the errors are zero on the outer barrel, where the same Dirichlet boundary condition is prescribed to all high-fidelity solutions. The same cannot be observed on the adiabatic screws -- particularly in regions where the warm melt sticks to the screw surface and is traced throughout the interior, the errors are the highest. Due to the right side being heated, the errors in the right barrel tend to be smaller than in the left.

At no point the temperature is off by more than $0.7$. The ratio of the typical range of errors to the typical range of temperature, i.e., $1.4/40$ happens to be close to the total standardized error for this sample at $\varepsilon_\text{POD-ANN}=0.034$, providing a good intuition behind the standardized error.

Lastly, we illustrate the effect of increasing the number of training samples $N_{tr}$. This is an easy to implement yet computationally expensive approach to reducing the errors. \reffig{fig:extr_400_errors} shows the errors over the number of bases $L$ using four times as many training samples $N_{tr}=400$ and the same validation and test sets as before.

For tuning the hyperparameters of the \glspl{ANN}, the upper bound for the search space of $N_{\nu}$ is increased to $100$ according to the heuristic discussed in \refsec{sec:ANN}. The regression errors for $\vekt{u}, p, T$ show the same qualitative behaviour as previously and the best identified $N_{\nu}$ and $\alpha_0$
are $50, 70, 70$ and $3{\times}10^{-3}, 1{\times}10^{-3}, 1{\times}10^{-3}$, respectively.

We observe that increasing $N_{tr}$ beyond $100$ has almost no effect on the projection error $\varepsilon_\text{POD}(\mathbb{P}_{te})$ suggesting that even less than $100$ samples can be used for the \gls{POD} with similar success.
However, the regression models benefit significantly from more data as the total standardized errors $\varepsilon_\text{POD-REG}(\mathbb{P}_{te})$ decrease around two-fold for \gls{RBF}, three-fold for \gls{ANN} and four- to eight-fold for \gls{GPR}. 
This shows another advantage of the \gls{GPR} while also demonstrating its biggest drawback, namely, the cubic complexity in data size $\mathcal{O}(N_{tr}^3)$ due to the inversion of the covariance matrix (see \refsec{sec:GPR}). The \gls{RBF} is also $\mathcal{O}(N_{tr}^3)$ due to the inversion of the weight matrix (see \refsec{sec:RBF}).
Meanwhile, \glspl{ANN} are known to scale very well with large datasets in practice, despite there not being a compact theoretical result for their time complexity \cite{Livni2014,boob2018,goel2017}. 


\FloatBarrier

\def\W{0.42\textwidth}
\begin{figure}[!htbp]
\begin{center}
\begin{tabular*}{\textwidth}{c @{\extracolsep{\fill}} c}
  \includegraphics[width=\W]{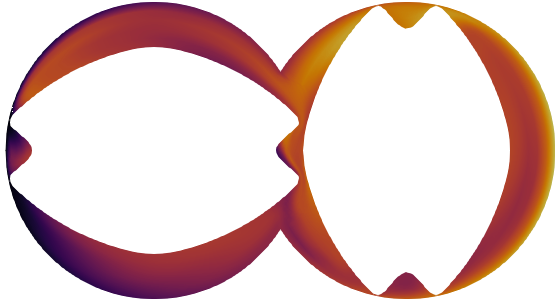} &
  \includegraphics[width=\W]{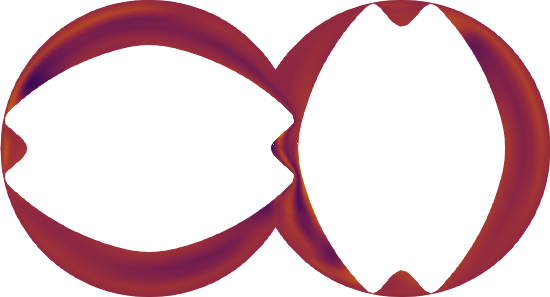} 
  \\
  \includegraphics[width=\W]{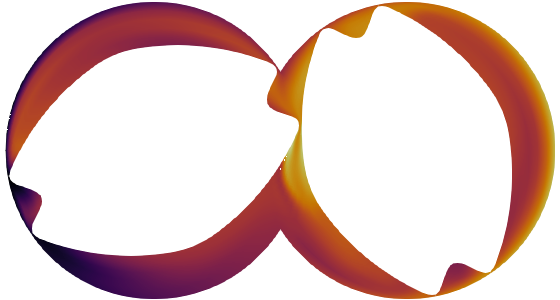} &
  \includegraphics[width=\W]{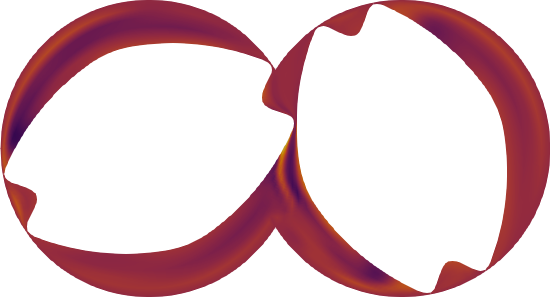}
  \\
  \includegraphics[width=\W]{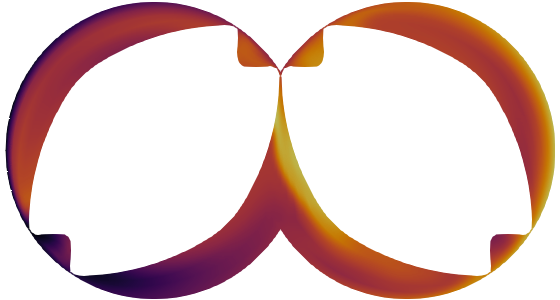} &
  \includegraphics[width=\W]{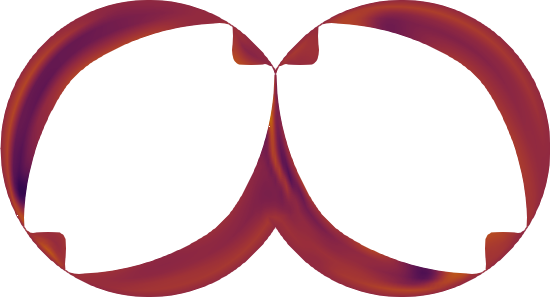}
  \\
  \includegraphics[width=\W]{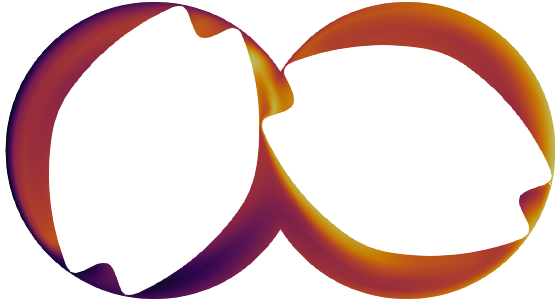} &
  \includegraphics[width=\W]{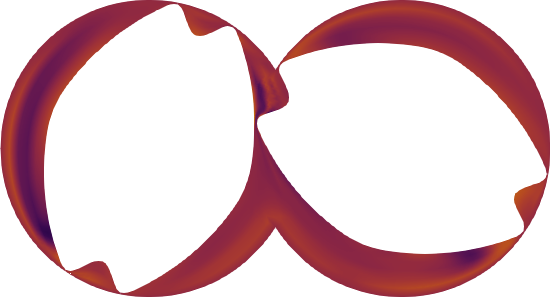}
  \\
  \includegraphics[width=\W]{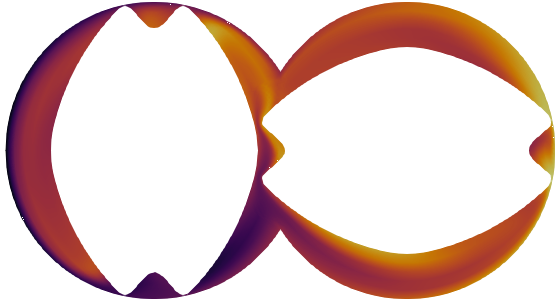} &
  \includegraphics[width=\W]{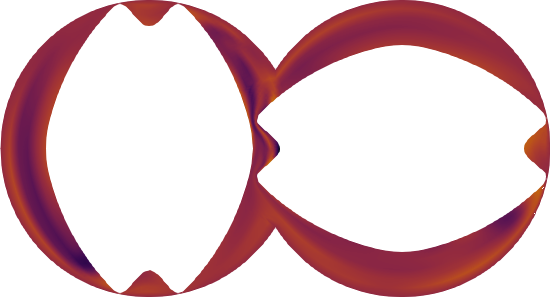} 
  \\
  \scaleNamedInferno{$454$}{$493$}{$T$} &
  \scaleNamedInferno{$-0.7$}{$0.7$}{$T-\tilde{T}$}
\end{tabular*}
\end{center}
\caption[True temperature distribution and error in prediction over several time steps for twin-screw extruder.]
{The high-fidelity true solution $T(\vekt{\mu}=(1.01,\allowbreak 6.52{\times}10^{12},\allowbreak 504) \in \mathbb{P}_{te})$ (left) and the error in prediction made by the best \gls{ANN} (right) at time steps $t_{60}$, $t_{65}$, $t_{70}$, $t_{75}$ and $t_{80}$. Over all investigated timesteps the standardized error of the prediction is $\varepsilon_\text{POD-ANN}=0.034$.
}
\label{fig:extr_truth_pred}
\end{figure}

\section{Conclusion}

In this work, we present a non-intrusive \gls{ROM} and apply it to three complex flow problems. 
The systems we study are characterized by unsteady non-linear parametrized \glspl{PDE} although the non-intrusive \gls{ROM} can be applied to general simulated or experimentally obtained data.
Our method extends the \gls{POD}-regression framework in several steps.
First, standardization of data removes scale and translation effects and allows to reuse similar regression model hyperparameters on different problems.
Second, our standardized error measure relates the error in the prediction to the variance in the dataset.
This makes the error interpretable even across problems of different scales and complexities.
Lastly, to deal with time-resolved data, we propose to treat time as a discrete dimension rather than a continuous parameter.
This preserves the efficiency of \gls{POD} and greatly simplifies the regression maps.

Our framework is first validated on a steady skewed lid-driven cavity problem. The results are in close agreement with those reported in the literature.
Next, we consider an unsteady oscillating lid-driven cavity problem.
Finally, we study a cross-section of a twin-screw extruder, which is characterized by a time- and temperature-dependent flow of a generalized Newtonian fluid on a moving domain. We vary the thermal conductivity, reference viscosity and density over an order of magnitude each.
Using only $100$ training samples, we obtain errors less than $3\%$ for the predicted velocity, pressure and temperature distributions. Although this already is within engineering bounds, the errors can be reduced significantly by simply generating more data. Using $400$ training samples, we achieve errors below $0.5\%$.

Additionally, we compare the performance of three different regression models: \gls{RBF} regression, \gls{GPR}, and \glspl{ANN}. 
The results of \gls{ANN}’s hyperparameter tuning suggests that standardization can alleviate this computationally expensive procedure and offer a good starting point for other practitioners -- in all results with 100 training samples, the optimal initial learning rate is between $10^{-3}$ and $10^{-2}$ and the width of the two hidden layers is between $20$ and $40$. Even so, \gls{GPR} is found to be a very competitive alternative to the \gls{ANN}, while being easier to train, interpret and control.
Especially in the strongly anisotropic and densely sampled oscillating lid-driven cavity problem, \gls{GPR} significantly outperforms the \gls{ANN}.
The \gls{RBF} regression consistently performs worse, but due to its simplicity it can serve as an inexpensive baseline.

Our standardization steps intend to facilitate a data-driven approach, but problem-specific adjustments to the method could still further boost the performance.
Examples include anisotropic or stabilized \gls{RBF} regression methods, more advanced or carefully chosen (e.g. non-stationary or periodic) \gls{GPR} kernels or deep-learning methods for \glspl{ANN}. However, if not in a data-sparse setting, the computational cost of the more advanced models should be weighted against simply generating more training data.

The implementation and the lid-driven cavity dataset are open-sourced\footnote{\url{https://github.com/arturs-berzins/sniROM}} and intend to serve as a turnkey baseline for other practitioners and facilitate the use of \gls{ROM} in general engineering workflows.

\section*{Acknowledgements}

The computations were conducted on computing clusters supplied by the IT Center of the RWTH Aachen University.

\FloatBarrier

\biboptions{sort&compress} 
\bibliographystyle{elsarticle-num}
\bibliography{Bibliography}

\end{document}